\documentclass[a4paper, 12pt]{article}

\usepackage{tikz}
\usepackage{latexsym,amsmath,amsfonts,amssymb}
\usepackage{amsthm}
\usepackage{mathtools}
\usepackage[latin1]{inputenc}
\usepackage[american]{babel}
\usepackage{bbm}
\usepackage[nosort]{cite}
\usepackage[colorlinks=true, citecolor=blue, linkcolor=blue, linktocpage=true]{hyperref}

\usepackage[symbol]{footmisc}
\usepackage{mathrsfs}

\renewcommand{\thefootnote}{\fnsymbol{footnote}}

\numberwithin{equation}{section}

\newcommand{\be}{\begin{equation}} \newcommand{\ee}{\end{equation}}
\newcommand{\bea}{\begin{equation} \begin{aligned}} \newcommand{\eea}{\end{aligned} \end{equation}}
\newcommand{\ba}{\begin{array}} \newcommand{\ea}{\end{array}}

\makeatletter
\def\blfootnote{\gdef\@thefnmark{}\@footnotetext}
\makeatother

\begin{document}

\thispagestyle{empty}

\vspace{12mm}
\begin{center}
{\large\bf Hawking Radiation,  Entanglement Entropy, \\ and Information Paradox of Kerr Black Holes}
\\[15mm]
{Jun Nian$^{1}$}
 
\bigskip
{\it
$^1$ International Centre for Theoretical Physics Asia-Pacific,\\ University of Chinese Academy of Sciences, 100190 Beijing, China\\[.5em]
}

{\tt nianjun@ucas.ac.cn}

\bigskip
\bigskip

{\bf Abstract}\\[5mm]
{\parbox{14cm}{\hspace{5mm}
The black hole information paradox is a long-standing problem in theoretical physics.  Despite some recent progress,  many issues remain open and should be clarified.  In this paper,  we study the information paradox of Kerr black holes and propose a new resolution with precise physical meanings.  We compute the time-dependent Hawking radiation rate during the Kerr black hole evaporation using both the gravity and the conformal field theory approaches.  Based on the consistent result from both approaches,  we formulate the information paradox on top of the time evolution of the entanglement entropy between a Kerr black hole and its Hawking radiation quanta.  To resolve the information paradox,  we carefully keep track of the ingoing Hawking quanta through the Kerr black hole as a quantum wormhole and microscopically derive the Page curve as a time-delay effect.  The result matches the previously obtained semi-classical Page curve and has a natural interpretation in quantum information theory.}}

\end{center}

\newpage
\pagenumbering{arabic}
\setcounter{page}{1}
\setcounter{footnote}{0}
\renewcommand{\thefootnote}{\arabic{footnote}}

\tableofcontents


\section{Introduction}

The black hole information paradox is probably the most famous problem in the field of gravitation.  It is believed that the resolution of this problem should pave the way towards the quantum theory of gravity.  Almost fifty years after this problem was raised,  there has been a lot of progress and a much deeper understanding,  but it still remains open to some extent.

Starting from Hawking's seminal work \cite{Hawking:1974sw},  it has been widely accepted that a black hole can radiate particles at a certain temperature,  which is called Hawking radiation.  Because these Hawking quanta carry energy away,  a black hole can slowly evaporate.  At the same time,  the black hole becomes quantum entangled with its Hawking radiation particles (also called Hawking quanta).  During the black hole evaporation,  the entanglement entropy seems to increase monotonically.  However,  when a black hole completely evaporates,  what can the previously radiated Hawking quanta entangle with? Is the quantum entanglement (or, in other words,  information) lost? This is the origin of the so-called information (loss) paradox.  During the last few decades,  many attempts have been made to resolve this paradox.  A very incomplete list of references includes \cite{Giddings:1992hh,  Hartle:1996rp,  Maldacena:2001kr,  Lunin:2001jy,  Horowitz:2003he,  Mathur:2005zp,  Skenderis:2008qn,  Almheiri:2012rt,  Papadodimas:2012aq,  Almheiri:2013hfa,  Papadodimas:2013wnh,  Papadodimas:2013jku,  Bradler:2013gqa,  Hawking:2016msc,  Penington:2019npb,  Almheiri:2019psf,  Penington:2019kki,  Almheiri:2019qdq}.  For some reviews see \cite{Preskill:1992tc,  Giddings:1995gd,  Mathur:2009hf,  Ashtekar:2020ifw,  Almheiri:2020cfm}.

To address this problem,  an enlightening perspective was provided by Page in \cite{Page:1993wv,  Page:2013dx}.   Viewing the black hole and its Hawking quanta as two subsystems,  he argued that the entanglement entropy between a black hole and its Hawking quanta should be dominated by the entropy of the smaller subsystem.  More precisely,  the entanglement entropy can be approximated by Hawking quanta's entropy at the beginning of evaporation but is roughly equal to the black hole entropy towards the end of evaporation.  As a function of time,  the entanglement entropy first increases from zero,  and after reaching a maximum,  it decreases again to zero.  This time evolution of entanglement entropy is called the Page curve.  How to microscopically derive the Page curve plays a central role in studying information paradox.

In recent years,  the Page curve has been derived for Schwarzschild black holes in Euclidean gravity \cite{Penington:2019npb,  Almheiri:2019psf,  Penington:2019kki,  Almheiri:2019qdq},  which is a milestone in the research field of black hole information paradox.  In this approach,  some new concepts have been introduced,  such as replica wormholes and the island formula,  which help deepen our understanding of the problem.  However,  this is probably not the end of the story because some issues still need to be clarified,  for instance,  a clear physical meaning of the island.

In this paper,  we will look at the black hole information paradox from a new perspective by adding a new ingredient: rotation.  The uncharged rotating black hole solution was first constructed by Kerr \cite{Kerr:1963ud},  and this class of black hole solutions is known as Kerr black holes.  We will see that the introduction of rotation significantly deforms the spacetime structure,  which makes the information paradox more tractable.

Kerr black holes play important roles in both observational and theoretical points of view.  Since the first discovery of gravitational waves from black holes in 2016 \cite{LIGOScientific:2016aoc}, much data on black holes has been obtained.  So far,  the gravitational wave detections show that the observed black holes are exclusively Kerr black holes.  Additionally,  the near-horizon geometry of Kerr black holes is intimately related to conformal field theory (CFT),  and various technical tools have been developed in this direction.

The main idea of this paper is to treat a Kerr black hole as a quantum wormhole,  which is already hinted at from the Penrose diagram of a non-extremal Kerr black hole.  With this in mind,  we compute the Hawking radiation rate using both the gravity and the CFT methods.  Based on the consistent result from both approaches,  we carefully keep track of a pair of maximally entangled Hawking radiation particles created at the outer horizon of a Kerr black hole.  The outgoing particle travels directly to the future infinity.  When the ingoing particle travels from the outer horizon to the inner horizon,  it is absorbed by the inner horizon. After a time delay, it is re-emitted from the white hole side towards the future infinity.  The entanglement is purified again when both particles arrive at the future infinity with a time difference.  However,  this time delay induces an evolution of entanglement entropy,  which increases at early time but decreases at late time, precisely reproducing the semi-classical result of a Kerr black hole's Page curve \cite{Nian:2019buz}.  The crucial step is to compute the time delay between these two maximally entangled particles,  where we will adopt some techniques from atomic physics.  We will also see that this step has a natural interpretation as quantum teleportation in quantum information theory.

This paper is organized as follows.  In Sec.~\ref{sec:QualitativePicture},  after reviewing some facts about Kerr black holes,  we briefly sketch the qualitative picture for the Hawking radiation and the evaporation of a Kerr black hole,  establishing a framework for a more detailed study.  In Sec.~\ref{sec:EEoldApproach},  we review the computation of Hawking radiation rate from the gravity side and reinterpret it in a CFT approach.  Based on the consistent results,  we compute the entanglement entropy from Hawking's perspective and demonstrate the information paradox in this context.  In Sec.~\ref{sec:PageCurve},  we first review the semi-classical computation of the Page curve and then derive it microscopically based on a time-delay effect between the maximally entangled Hawking quanta.  We also show that the crucial absorption and emission process at the inner horizon has a natural quantum information interpretation.  Finally, in Sec.~\ref{sec:Discussion},  we discuss some possible future directions and the connection between our new approach and the previous works in literature.  Throughout this paper,  we use the unit convention with $G_N = \hbar = k_B = 1$.

\section{Qualitative Picture of Kerr Black Hole Evaporation}\label{sec:QualitativePicture}

In this section,  we first review some known facts on Kerr black holes in the literature.  After that,  we will provide a qualitative picture of Kerr black hole evaporation and discuss how the information paradox can potentially be resolved in this context.  More quantitative details will be presented in the following sections.

\subsection{Review of Kerr Black Holes}

In the Boyer-Lindquist coordinates,  the metric of a Kerr black hole is
\begin{align}
  ds^2 & = - \left(1 - \frac{2 M r}{\rho^2} \right)\, dt^2 + \left(r^2 + a^2 + \frac{2 a^2 M r\, \textrm{sin}^2 \theta}{\rho^2} \right)\, \textrm{sin}^2 \theta\, d\phi^2 - \frac{4 a M r\, \textrm{sin}^2 \theta}{\rho^2}\, d\phi\, dt \nonumber\\
  {} & \quad + \frac{\rho^2}{\Delta}\, dr^2 + \rho^2\, d\theta^2\, ,\label{eq:KerrMetric}
\end{align}
where $a \equiv J / M$ with $J$ and $M$ denoting the mass and the angular momentum of the black hole,  respectively,  and
\be
  \rho^2 \equiv r^2 + a^2\, \textrm{cos}^2 \theta\, ,\quad \Delta \equiv r^2 + a^2 - 2 M r\, .
\ee
The Kerr black hole has an outer horizon and an inner horizon located at $r_+$ and $r_-$,  respectively,  with
\be
  r_\pm \equiv M \pm \sqrt{M^2 - a^2}\, .
\ee
The Hawking temperature $T_H$ and the angular velocity $\Omega$ can be defined as
\be
  T_H \equiv \frac{r_+ - r_-}{8 \pi M r_+}\, ,\quad \Omega \equiv \frac{a}{2 M r_+}\, ,
\ee
while the Bekenstein-Hawking entropy is given by
\be\label{eq:S_BH}
  S = 2 \pi (M^2 + \sqrt{M^4 - J^2})\, .
\ee
With these quantities,  the first law of black hole thermodynamics can be written as
\be
  dM = T_H\, dS + \Omega\, dJ\, .
\ee
When
\be\label{eq:ExtremalLimit}
  a = M\, , \quad \text{or equivalently}\quad J = M^2\, ,
\ee
the outer and the inner horizons coincide,  consequently $T_H = 0$.  The Kerr black hole obeying this condition \eqref{eq:ExtremalLimit} is called an extremal Kerr black hole.

\begin{figure}[!htb]
      \begin{center}
        \includegraphics[width=0.37\textwidth]{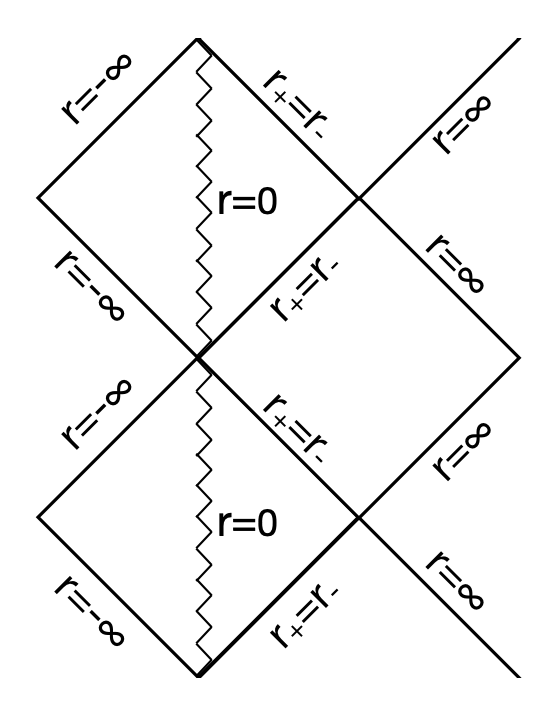}
        \caption{The Penrose diagram of an extremal Kerr black hole}
        \label{fig:KerrBHPenroseDiagramExt}
      \end{center}
    \end{figure}

\begin{figure}[!htb]
      \begin{center}
        \includegraphics[width=0.33\textwidth]{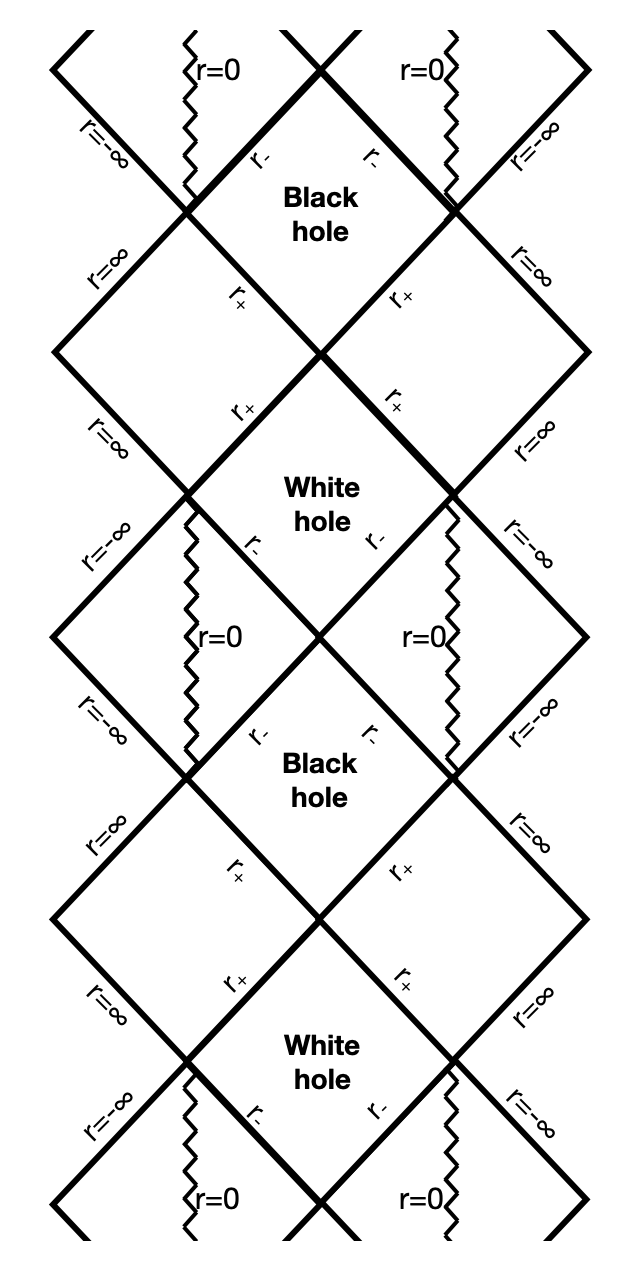}
        \caption{The Penrose diagram of a non-extremal Kerr black hole}
        \label{fig:KerrBHPenroseDiagramNonExt}
      \end{center}
    \end{figure}

The spacetime structure of an eternal extremal Kerr black hole and a generic eternal non-extremal Kerr black hole can be characterized by their Penrose diagrams in Figs.~\ref{fig:KerrBHPenroseDiagramExt} and \ref{fig:KerrBHPenroseDiagramNonExt},  respectively.

\newpage

\subsection{Kerr Black Hole as Wormhole}\label{sec:KerrWormhole}

The wormhole as a possible solution in general relativity has been extensively studied in the literature,  initiated by the celebrated works \cite{MorrisThorne1988,  MorrisThorneYurtsever1988}.  Most of the previously studied wormhole solutions are spacelike wormholes.

Observing the Penrose diagrams of Kerr black holes shown in the previous subsection,  we see that the maximally extended spacetime of a Kerr black hole contains both black hole and white hole parts,  which can be effectively viewed as a wormhole.  Let us call it a Kerr wormhole.  Following a timelike curve,  an object can potentially travel through this wormhole to another patch of the universe.  Hence,  the Kerr wormhole is timelike.  However,  there are some issues with the traversability of the Kerr wormhole,  such as the inner horizon's stability as a Cauchy horizon and the violation of strong cosmic censorship,  which we will discuss later in the paper.

Nevertheless,  even if a Kerr wormhole cannot be used for interstellar travel,  this novel perspective can possibly provide a new way for helping resolve the long-standing information paradox of black holes.  The main idea is to carefully keep track of the Hawking radiation quanta in the interior of the Kerr wormhole,  and then compute the time evolution of the entanglement entropy between the Kerr wormhole and the Hawking quanta.

From the Penrose diagrams of Kerr black holes (Figs.~\ref{fig:KerrBHPenroseDiagramExt} and \ref{fig:KerrBHPenroseDiagramNonExt}),  we see that the maximally extended universe can be generically divided into two different types of regions:
\begin{itemize}
  \item The region outside the outer horizons with $r > r_+$,  such as our universe;
  
  \item The region inside the outer horizons with $r < r_+$.  Mathematically,  this type of region can be extended to $r \to - \infty$.  However, we only consider $0 \leq r < r_+$ in physics.
\end{itemize}

A timelike or lightlike particle can cross the borders of different regions and travel from one patch to another.  Although different patches are sometimes viewed as different universes in the literature,  we will treat them as separate patches of the same universe connected through the Kerr black holes (including white hole sides).  In addition,  we identify the past infinities $\mathscr{I}^-$,  the future infinities $\mathscr{I}^+$,  and the spatial infinities ${\it i}_0$ from different patches,  respectively.  Hence,  the Kerr black hole indeed plays the same role as a wormhole in the sense that it interconnects different patches of the universe.

\begin{figure}[!htb]
      \begin{center}
        \includegraphics[width=0.45\textwidth]{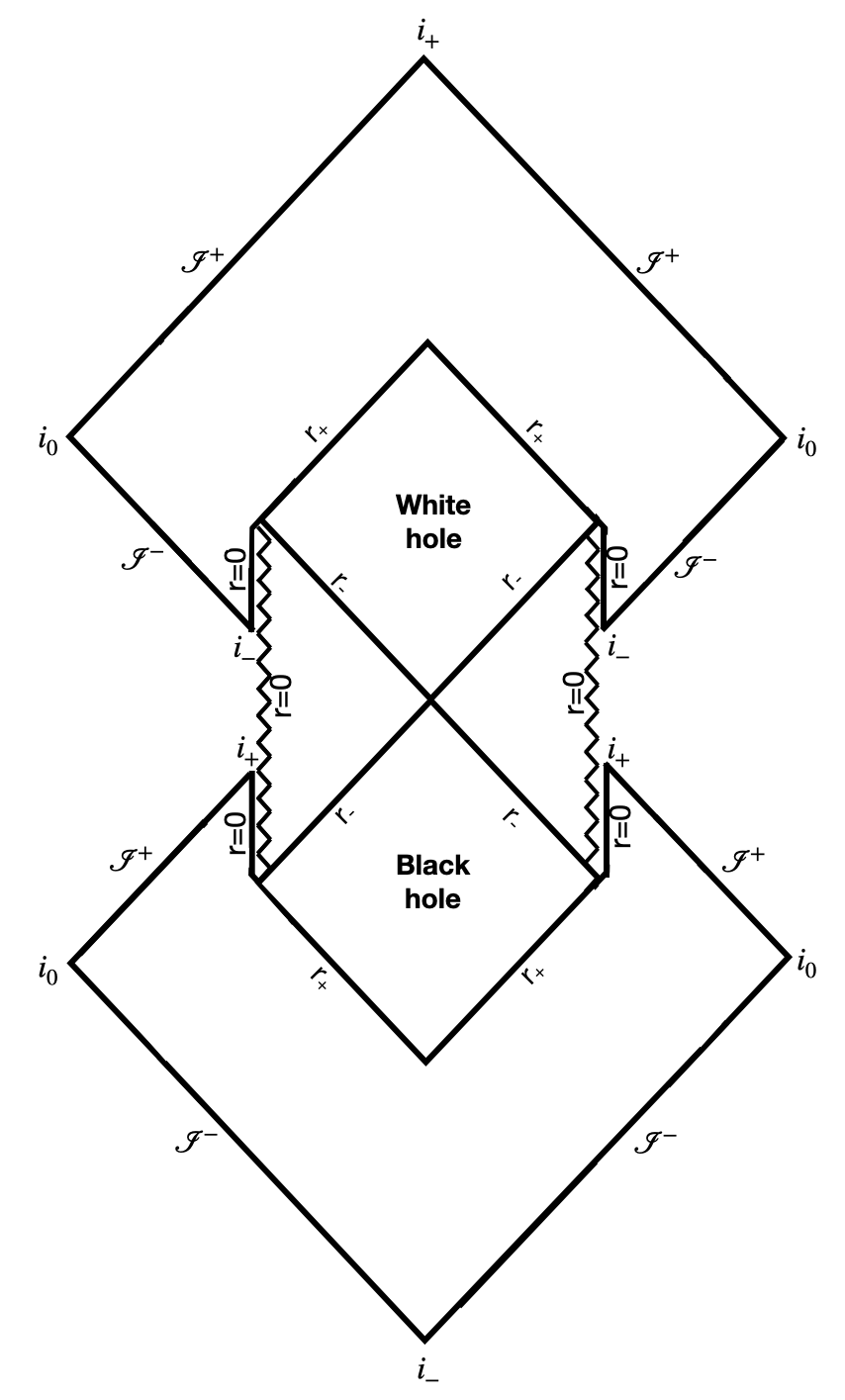}
        \caption{The Penrose diagram of a Kerr black hole from formation to evaporation}
        \label{fig:EvaporatingKerrBHPenroseDiagram}
      \end{center}
    \end{figure}

\begin{figure}[!htb]
      \begin{center}
        \includegraphics[width=0.41\textwidth]{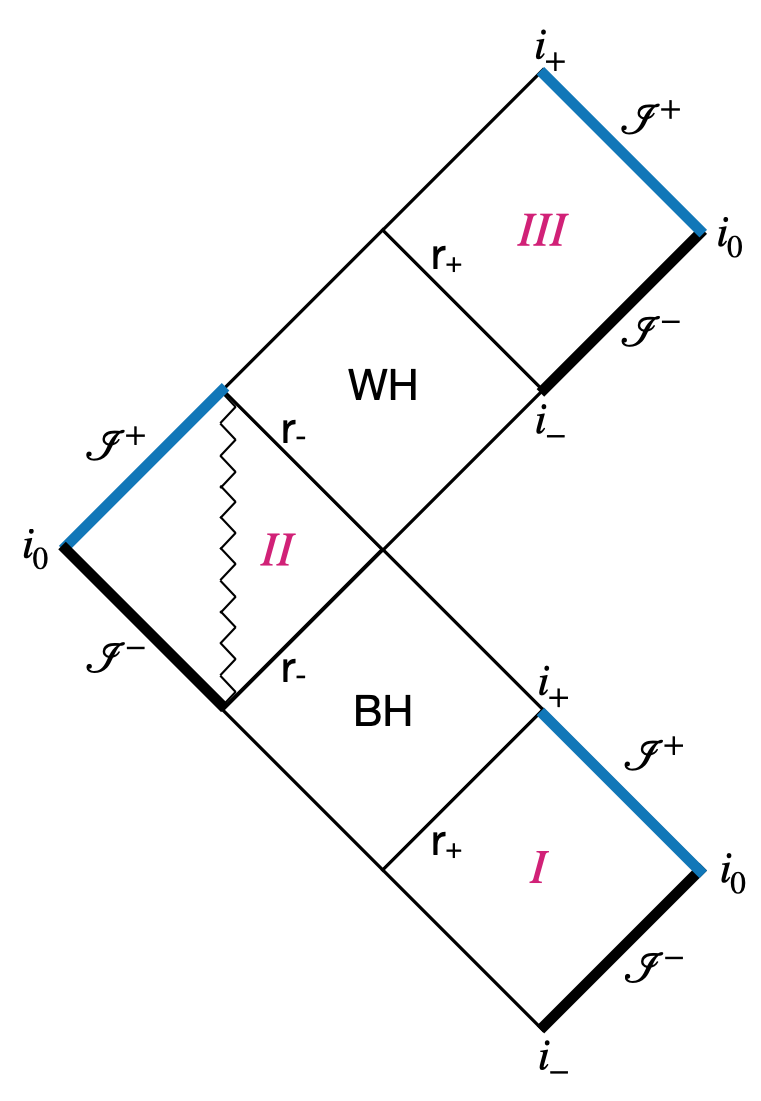}
        \caption{A non-extremal Kerr black hole's Penrose diagram with three representative regions}
        \label{fig:KerrBHPenroseDiagram3Regions}
      \end{center}
    \end{figure}

In principle,  the Penrose diagrams of eternal Kerr black holes shown in Figs.~\ref{fig:KerrBHPenroseDiagramExt} and \ref{fig:KerrBHPenroseDiagramNonExt} are infinitely extended and contain infinitely many separate patches of the universe.  A more realistic Kerr black hole should be formed from collapsing matter and eventually completely evaporate, schematically shown in Fig.~\ref{fig:EvaporatingKerrBHPenroseDiagram}.  Before the black hole is formed and after the black hole has completely evaporated,  the points at $r=0$ are nonsingular.  In this paper,  to simplify the discussion of the information paradox during the Kerr black hole evaporation,  we will focus on several representative regions shown in Fig.~\ref{fig:KerrBHPenroseDiagram3Regions},  which describe the spacetime structure after a Kerr black hole is formed and before it is completely evaporated.

Treating the three regions in Fig.~\ref{fig:KerrBHPenroseDiagram3Regions} as separate regions of the same universe,  we conjecture for the corresponding quantum theory that the full Hilbert space consists of three subspaces,  i.e.,
\be
  \mathcal{H}_{\text{full}} = \mathcal{H}_I \otimes \mathcal{H}_{II} \otimes \mathcal{H}_{III}\, ,
\ee
where the three regions are labeled by $I$,  $II$,  and $III$,  respectively (see Fig.~\ref{fig:KerrBHPenroseDiagram3Regions}),  and Region $II$ covers the spacetime within $0 \leq r < r_+$ including both the black hole and the white hole parts.

The standard Hawking radiation takes place in the vicinity of the outer horizon $r_+$ in Region $I$.  A pair of maximally entangled lightlike particles $1$ and $2$ are created from the vacuum.  The particle $1$ will move towards the future infinity $\mathscr{I}^+$ in Region $I$,  while the particle $2$ enters the outer $r_+$ horizon and after some time arrives at the inner horizon $r_-$.  According to the previous analyses \cite{Hawking:1974sw,  Wald:1995yp},  the particle $2$ carries negative energy.  After the particle $2$ enters the outer horizon,  it becomes part of the black hole,  or in other words,  it is eaten by the black hole and causes the decrease of the black hole mass.  At the same time,  the particle $2$ induces the entanglement between the black hole and the Hawking particle $1$.

Repeating this procedure,  a vast number of such Hawking particle pairs are created and lead to black hole evaporation.  In the traditional picture of the Schwarzschild black hole evaporation,  the particles entering the black hole have no way out and are trapped within the event horizon.  Correspondingly,  the entanglement between the black hole and the outgoing Hawking quanta increases monotonically,  which is incompatible with the complete evaporation of the black hole in the final state.  This is called the black hole information paradox.  There have been many attempts in the literature trying to resolve this paradox.  A very incomplete list includes,  e.g.,  \cite{Giddings:1992hh,  Hartle:1996rp,  Maldacena:2001kr,  Lunin:2001jy,  Horowitz:2003he,  Mathur:2005zp,  Skenderis:2008qn,  Almheiri:2012rt,  Papadodimas:2012aq,  Almheiri:2013hfa,  Papadodimas:2013wnh,  Papadodimas:2013jku,  Bradler:2013gqa,  Hawking:2016msc,  Penington:2019npb,  Almheiri:2019psf,  Penington:2019kki,  Almheiri:2019qdq}.

As mentioned above,  we have identified in different patches the past infinities $\mathscr{I}^-$ and the future infinities $\mathscr{I}^+$,  respectively.  Consequently,  an observer at the future infinity $\mathscr{I}^+$ can collect Hawking quanta arrived at $\mathscr{I}^+$ from different patches.


\subsection{Inside Inner Horizon}\label{sec:InnerHorizon}

\begin{figure}[!htb]
      \begin{center}
        \includegraphics[width=0.45\textwidth]{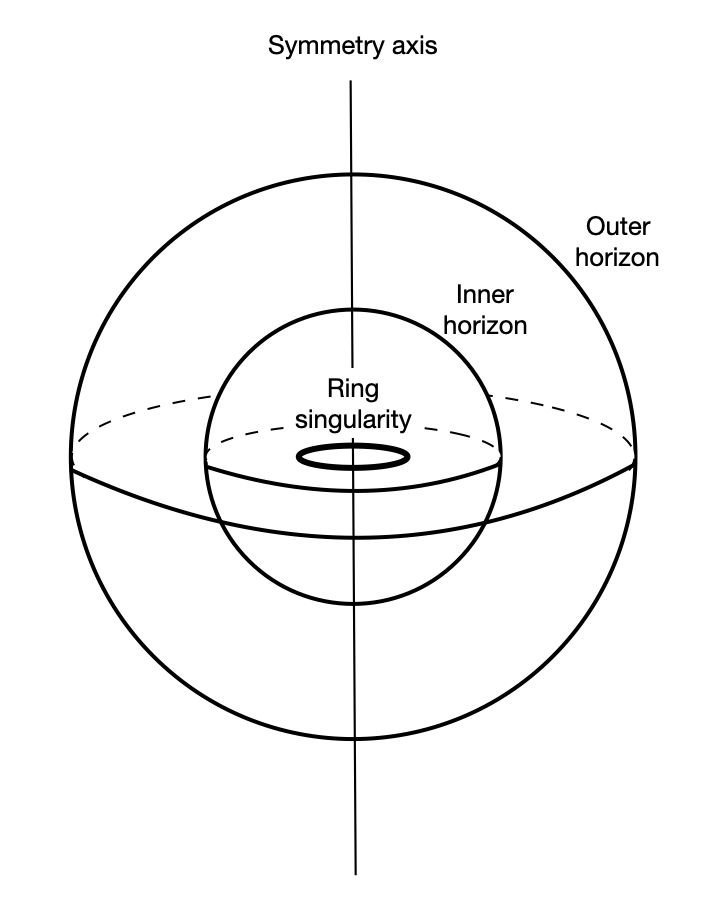}
        \caption{A sketch of the spatial structure of a Kerr black hole's interior \cite{Hawking:1973uf}}
        \label{fig:KerrBHSketch}
      \end{center}
    \end{figure}

It is well known in the literature that a Kerr black hole has an outer horizon and an inner horizon.  In contrast to a Schwarzschild black hole, which has a spacelike singularity,  a Kerr black hole has a timelike singularity,  which has the shape of a ring in spatial dimensions (see Fig.~\ref{fig:KerrBHSketch}).

What happens when a lightlike particle enters the outer horizon of a Kerr black hole? According to general relativity, nothing special should happen until it arrives at the inner horizon.  After that,  there is no definite answer in general relativity because the inner horizon is a Cauchy horizon,  and a lightlike particle passing through the inner horizon may violate the strong cosmic censorship.  We will comment on this issue later in Sec.~\ref{sec:Discussion}.

In this paper,  we will treat the inner horizon and its interior as an atom,  generalizing some similar ideas previously studied in the literature \cite{Bekenstein:1997bt,  EmparanSachs}.  More precisely,  we assume that the inner horizon and its interior absorb the incoming particle $2$ and re-emit a new particle $2'$ from the white hole side after a time delay.  This new particle $2'$ maintains the maximal entanglement with the outgoing particle $1$ created at the black hole's outer horizon.  Hence,  talking about the entanglement between Hawking quanta and the black hole-white hole system is more accurate than only the black hole.  We will see that this absorption and emission process is fully quantum-mechanical and has an interpretation in quantum information theory,  which releases the tension with the strong cosmic censorship.

After the new particle $2'$ is emitted from the inner horizon on the white hole side,  it can pass through the outer horizon of the white hole and eventually arrive at the future infinity $\mathscr{I}^+$,  where the observer can collect particles in a maximally entangled pair from two patches.  Consequently,  the entanglement between the black hole-white hole system and Hawking quanta will be purified,  and no information paradox is present.  However,  the particle $2'$ does not arrive at $\mathscr{I}^+$ simultaneously with $1$.  Instead,  it is time-delayed compared to the particle $1$.  This time-delay effect will cause the entanglement entropy between the black hole-white hole system and Hawking quanta to first increase to a maximum and then decrease to zero,  which is known as the Page curve \cite{Page:1993wv,  Page:2013dx,  Nian:2019buz}.  The main goals of this paper are to describe this mechanism in detail and microscopically compute the Page curve for a Kerr black hole.

\section{Entanglement Entropy in Kerr Black Hole Evaporation}\label{sec:EEoldApproach}

\subsection{Hawking Radiation Rate from Gravity Side}\label{eq:RateViaGravity}

The Hawking radiation rate of a Schwarzschild black hole was computed by Hawking himself in a seminal paper \cite{Hawking:1974sw}.  The radiation rate for a rotating black hole has been calculated by solving a wave equation of a spin-$s$ field in the rotating black hole background with appropriate boundary conditions.  This kind of equation was first systematically studied by Teukolsky \cite{Teukolsky1972}.  Using this approach,  the Hawking radiation rates of rotating black holes were first computed in \cite{Starobinsky:1973aij,  Starobinsky:1974nkd} by Starobinsky and Churilov for integer-spin fields and later generalized by Teukolsky and Press for the extremal and the superradiant cases \cite{Teukolsky:1974yv} and by Page for half-integer spins \cite{Page1976}.  In this subsection,  we briefly review the computation of the Hawking radiation rate from the gravity side,  whose CFT interpretation will be discussed in Sec.~\ref{sec:CFTinterpretation}.

Consider a spin-$s$ massless field of frequency $\omega$ and angular momentum quantum number $m$ in a Kerr black hole background.  Using the Boyer-Lindquist coordinates for the Kerr black hole and the Ansatz
\be
  \Psi = e^{- i \omega t + i m \phi}\, R(r)\, S (\theta)\, ,
\ee
we can express the wave equation
\be\label{eq:KerrWaveEq}
  \partial_\mu \left(\sqrt{- g}\, g^{\mu\nu} \partial_\nu \Psi \right) = 0
\ee
into the angular and the radial parts.  In terms of the following new quantities:
\begin{align}
\begin{split}\label{eq:Definitions}
  x & \equiv \frac{r - r_+}{2 (r_+ - M)} = \frac{r - M - \sqrt{M^2 - a^2}}{2 \sqrt{M^2 - a^2}}\, ,\\
  Q & \equiv \frac{m \Omega - \omega}{2 \kappa} = \frac{M r_+}{r_+ - M} (m \Omega - \omega)\, ,\\
  A & \equiv 8 \pi M r_+ = 4 \pi (r_+^2 + a^2) = 8 \pi M (M + \sqrt{M^2 - a^2})\, ,\\
  \kappa & \equiv \frac{4 \pi (r_+ - M)}{A}\, ,\quad k \equiv 2 \omega (r_+ - M) = \frac{\kappa \omega A}{2 \pi}\, ,
\end{split}
\end{align}
the angular and the radial equations in the low-energy limit $M \omega \ll 1$ (also implying $k \ll 1$) are
\begin{align}
  \frac{1}{\textrm{sin}\, \theta} \frac{d}{d\theta} \left(\textrm{sin}\, \theta\, \frac{dS}{d\theta} \right) + \left[\left(s - a \omega\, \textrm{cos}\, \theta \right)^2 - \left(\frac{m + s\, \textrm{cos}\, \theta}{\textrm{sin}\, \theta} \right)^2 - s (s - 1) + \lambda - a^2 \omega^2 \right] S & = 0\, ,\\
  x^2 (x+1)^2\, \frac{d^2 R}{dx^2} + (s+1) x (x+1) (2 x + 1)\, \frac{dR}{dx} \qquad\qquad\qquad\qquad\qquad\qquad\qquad\qquad & {} \nonumber\\
  + \left[k^2 x^4 + 2 i s k x^3 - \lambda x (x + 1) + i s Q (2 x + 1) + Q^2 \right] R & = 0\, ,\label{eq:RadialWaveEqLowEnergy}
\end{align}
where the separation constant is $\lambda = (l - s) (l + s + 1)$ in the low-energy limit,  and $l$ denotes the spherical harmonic quantum number.

Let us focus on the radial wave equation \eqref{eq:RadialWaveEqLowEnergy}.  It can be solved in the near-horizon region $k x \ll l + 1$.  A solution with the ingoing boundary condition at the horizon is
\be\label{eq:NearSol}
  R = x^{-s + i Q}\, (x+1)^{-s - i Q} \phantom{|}_2 F_1 \left(- l - s,\, l - s + 1;\, 1 - s + 2 i Q;\, - x \right)\, .
\ee
In the asymptotic far region $x \gg |Q| + 1$,  the radial wave equation \eqref{eq:RadialWaveEqLowEnergy} can also be solved analytically:
\be\label{eq:FarSol}
  R = C_1\, e^{-i k x}\, x^{l - s}\phantom{|}_1 F_1 (l - s + 1;\, 2 l + 2;\, 2 i k x) + C_2\, e^{- i k x}\, x^{- l - s - 1}\phantom{|}_1 F_1 (- l - s;\, - 2 l;\, 2 i k x)\, .
\ee
Gluing the solutions \eqref{eq:NearSol} and \eqref{eq:FarSol} in the overlapping region $|Q| + 1 \ll x \ll (l+1)/k$,  we can fix the constants
\be
  C_1 = \frac{\Gamma (2 l + 1)\, \Gamma (1 - s + 2 i Q)}{\Gamma (l - s + 1)\, \Gamma (l + 1 + 2 i Q)}\, ,\quad C_2 = \frac{\Gamma (- 2 l - 1)\, \Gamma (1 - s + 2 i Q)}{\Gamma (- l - s)\, \Gamma (- l + 2 i Q)}\, .
\ee
With these constants $C_{1,2}$,  the solution \eqref{eq:FarSol} in the far region has the following asymptotic expansion for $k x \gg 1$:
\be\label{eq:AsymptSol}
  R = Y_{\text{in}}\, e^{- i k x}\, r^{-1} + Y_{\text{out}}\, e^{i k x}\, r^{- 2 s - 1}\, ,
\ee
where
\begin{align}
  Y_{\text{in}} & = \frac{\Gamma (2 l + 1)\, \Gamma (2 l + 2)\, \Gamma (1 - s + 2 i Q)}{\Gamma (l - s + 1)\, \Gamma (l + s + 1)\, \Gamma (l + 1 + 2 i Q)}\, \frac{k}{\omega}\, (- 2 i k)^{-l + s - 1} \\
  {} & \quad + \frac{\Gamma (- 2 l)\, \Gamma (- 2 l - 1)\, \Gamma (1 - s + 2 i Q)}{\Gamma (- l - s)\, \Gamma (- l + s)\, \Gamma (- l + 2 i Q)}\, \frac{k}{\omega}\, (-2 i k)^{l + s}\, ,\\
  Y_{\text{out}} & = \frac{\Gamma (2 l + 1)\, \Gamma (2 l + 2)\, \Gamma (1 - s + 2 i Q)}{\left[\Gamma (l - s + 1) \right]^2\, \Gamma (l + 1 + 2 i Q)}\, \left(\frac{k}{\omega} \right)^{2 s + 1} (2 i k)^{- l - s - 1} \\
  {} & \quad + \frac{\Gamma (- 2 l)\, \Gamma (- 2 l - 1)\, \Gamma (1 - s + 2 i Q)}{\left[\Gamma (- l - s) \right]^2\, \Gamma (- l + 2 i Q)}\, \left(\frac{k}{\omega}\right)^{2 s + 1} (2 i k)^{l - s}\, .
\end{align}
The first term in \eqref{eq:AsymptSol} plays the role of the ingoing wave at the asymptotic infinity,  while the second term stands for the outgoing wave at the asymptotic infinity,  which can be viewed as reflected by the Kerr black hole.  Suppose the absorption probability of the black hole measured at the asymptotic infinity is $\Gamma$.  Then,  the corresponding reflection probability is $1 - \Gamma$.  The emission probability of a black hole is equal to its absorption probability $\Gamma$.

For a spin-$0$ field,  the reflection probability can be computed as
\be
  1 - \Gamma = \bigg|\frac{Y_{\text{out}}}{Y_{\text{in}}} \bigg|^2\, ,
\ee
which does not hold for a generic spin-$s$ ($s \neq 0$) field.  To compute the reflection probability,  a method has been introduced in \cite{Teukolsky:1974yv}.  The asymptotic expansion \eqref{eq:AsymptSol} is for spin $s$.  When the spin is flipped,  i.e.,  $s \to - s$,  the corresponding asymptotic expansion is
\be
  R_{-s} = Z_{\text{in}}\, e^{- i k x}\, r^{-1} + Z_{\text{out}}\, e^{i k x}\, r^{2 s - 1}\, .
\ee
From the results of \cite{Teukolsky:1974yv},  we can deduce that the reflection probability for a generic spin-$s$ field is given by
\be
  1 - \Gamma = \bigg|\frac{Y_{\text{out}}}{Y_{\text{in}}}\cdot \frac{Z_{\text{out}}}{Z_{\text{in}}} \bigg|\, .
\ee
The low-energy limit implies that $k \ll 1$.  Using the above approach,  one can derive at the leading order in small $k$:
\be
  \Gamma = \text{Re} \Bigg[4\, e^{i \pi (s - \frac{1}{2})}\, \textrm{cos} \left(\pi (l - s) \right)\cdot \frac{\Gamma (- 2 l)\, \Gamma (- 2 l - 1)}{\Gamma (2 l + 1)\, \Gamma (2 l + 2)} \left(\frac{\Gamma (l - s + 1)}{\Gamma (- l - s)} \right)^2 \frac{\Gamma (l + 1 + 2 i Q)}{\Gamma (- l + 2 i Q)}\, (2 k)^{2 l + 1} \Bigg]\, ,
\ee
which can be further simplified under the assumptions $2 s \in \mathbb{Z}$ and $l - s \in \mathbb{Z}_{\geq 0}$.  Hence,  for an asymptotically flat Kerr black hole in (3+1) dimensions,  the absorption probability measured at the asymptotic infinity in the low-energy limit is
\begin{align}\label{eq:Gamma}
  \Gamma & = \text{Re} \Bigg[ \left(\frac{(l - s)!\, (l + s)!}{(2 l)!\, (2 l + 1)!!} \right)^2\, \frac{\Gamma (l + 1 + 2 i Q)}{\Gamma (- l + 2 i Q)}\, (2 i k)^{2 l + 1} \Bigg] \nonumber\\
  {} & = \left\{
  \begin{aligned}
    & \left(\frac{(l - s)!\, (l + s)!}{(2 l)!\, (2 l + 1)!!} \right)^2\, \prod_{n=1}^l \left[1 + \left(\frac{\omega - m \Omega}{n \kappa} \right)^2 \right]\, 2 \left(\frac{\omega - m \Omega}{\kappa} \right)\, \left(\frac{A \kappa}{2 \pi}\, \omega \right)^{2 l + 1}\, , \quad \textrm{for } s \in \mathbb{Z}\, ; \\
    & \left(\frac{(l - s)!\, (l + s)!}{(2 l)!\, (2 l + 1)!!} \right)^2\, \prod_{n=1}^{l + \frac{1}{2}} \left[1 + \left(\frac{\omega - m \Omega}{\left(n - \frac{1}{2} \right) \kappa} \right)^2 \right]\, \left(\frac{A \kappa}{2 \pi}\, \omega \right)^{2 l + 1}\, , \quad \textrm{for } s \notin \mathbb{Z} \text{ and } 2 s \in \mathbb{Z}\, ,
  \end{aligned} \right.
\end{align}
which can be interpreted as the emission probability of a Kerr black hole.

Using this result,  the particle rate of Hawking radiation for a specific particle can be expressed as
\be\label{eq:RateFromGravity}
  \langle N_{s \omega l m p} \rangle = \Gamma_{s \omega l m p}\, \frac{1}{\textrm{exp} \left[\frac{2 \pi}{\kappa} (\omega - m \Omega) \right] \mp 1}\, ,
\ee
where $\Gamma_{s \omega l m p}$ is given by $\Gamma$ in \eqref{eq:Gamma} for fixed quantum numbers $(s, \omega,  l,  m,  p)$,  and the choice of the sign in the denominator specifies the Bose-Einstein or the Fermi-Dirac statistics.  Consequently,  Page derived the following ordinary differential equation system of a Kerr black hole \cite{Page1976}:
\be\label{eq:ODE}
  - \frac{d}{dt} \left(\begin{array}{c}
  M \\
  J \end{array}\right)
  = \sum_{slmp} \frac{1}{2 \pi} \int d\omega\, \Gamma_{s \omega l m p} \frac{1}{\textrm{exp} \left[\frac{2 \pi}{\kappa} (\omega - m \Omega) \right] \mp 1}
  \left(\begin{array}{c}
  \omega \\
  m
  \end{array}\right)\, .
\ee
By solving this differential equation system,  we obtain $\left(M(t),\, J(t) \right)$ as functions of time for a Kerr black hole \cite{Nian:2019buz}.

For a fixed spin $s$,  the prefactor in the radiation probability \eqref{eq:Gamma} is maximal at $l = s$.  Hence,  the case $l = s$ provides the dominant contribution to the emission.  For instance,  for the emission of photons,  the dominant contribution to the Hawking radiation rate is
\begin{align}
  \langle N_{1 \omega 1 m p} \rangle & = \Gamma_{1 \omega 1 m p}\, \frac{1}{\textrm{exp} \left[\frac{1}{T_H} (\omega - m \Omega) \right] - 1} \nonumber\\
  {} & = \frac{2}{3} \left[1 + \left(\frac{\omega - m \Omega}{n \kappa} \right)^2 \right] \left(\frac{\omega - m \Omega}{\kappa} \right)\, \left(\frac{A \kappa}{2 \pi}\, \omega \right)^3 \frac{1}{\textrm{exp} \left[\frac{1}{T_H} (\omega - m \Omega) \right] - 1}\, .
\end{align}
Recall that for the (3+1)-dimensional blackbody radiation,  the particle number emission rate at a given frequency per unit area and per solid angle is
\be
  B (\omega,\, T) = \frac{\omega^2}{2 \pi^2 c^2}\, \frac{1}{\textrm{exp} \left(\frac{\hbar \omega}{k_B T} \right) - 1}\, .
\ee
By comparing these two expressions,  we see that the (3+1)-dimensional Hawking radiation rate deviates from the blackbody radiation and has a greybody factor
\be
  \sim \left[1 + \left(\frac{\omega - m \Omega}{n \kappa} \right)^2 \right] \left(\frac{\omega - m \Omega}{\kappa} \right) \left(\frac{A \kappa}{2 \pi}\, \omega \right)\, .
\ee

\subsection{Review of Near-Horizon Conformal Symmetry}\label{eq:KerrCFTReview}

To quantitatively compute the time evolution of the entanglement entropy between the Kerr black hole and the Hawking quanta,  i.e.,  the Page curve,  we must first calculate the Hawking radiation rate at a given time during the black hole evaporation.  This problem can be solved in gravity by solving the wave equation,  as shown in the previous subsection.  However,  it can also be converted via the Kerr/CFT correspondence into a transition rate problem in 2d conformal field theory,  and this picture provides some new insight.  In this subsection,  let us recall some known facts about the Kerr/CFT correspondence and its various extensions in the literature,  which are relevant to this paper.  For a more thorough review,  see \cite{Compere:2012jk} and the references therein.

{\flushleft $\bullet$ Extremal Kerr/CFT:}

The Kerr/CFT correspondence was first established for asymptotically flat extremal Kerr black holes \cite{Guica:2008mu}.  Under the extremality condition $a = M$,  we consider the near-horizon region of a Kerr black hole.  We use the following new coordinates:
\be\label{eq:NHscaling}
  \hat{t} \equiv \frac{\lambda\, t}{2 M}\, ,\quad \hat{y} \equiv \frac{\lambda M}{r - M}\, ,\quad \hat{\phi} \equiv \phi - \frac{t}{2 M}\, ,
\ee
which were first introduced by Bardeen and Horowitz \cite{Bardeen:1999px}.  Taking the limit $\lambda \to 0$ of the metric \eqref{eq:KerrMetric},  we obtain a near-horizon extreme Kerr (NHEK) geometry given by the metric:
\be\label{eq:NHEK}
  ds^2 = 2 J \widetilde{\Omega}^2 \left[\frac{- d\hat{t}^2 + d\hat{y}^2}{\hat{y}^2} + d\theta^2 + \widetilde{\Lambda}^2 \left(d\hat{\phi} + \frac{d\hat{t}}{\hat{y}} \right)^2 \right]\, ,
\ee
where
\be
  \widetilde{\Omega}^2 \equiv \frac{1 + \textrm{cos}^2 \theta}{2}\, ,\quad \widetilde{\Lambda} \equiv \frac{2\, \textrm{sin}\, \theta}{1 + \textrm{cos}^2 \theta}\, .
\ee
After a further change of coordinates:
\begin{align}
\begin{split}
  \hat{y} & = \left(\textrm{cos}\, \tau \sqrt{1 + \widetilde{r}^2} + \widetilde{r} \right)^{-1}\, ,\\
  \hat{t} & = \hat{y}\, \textrm{sin}\, \tau \sqrt{1 + \widetilde{r}^2}\, ,\\
  \hat{\phi} & = \varphi + \textrm{log} \left(\frac{\textrm{cos}\, \tau + \widetilde{r}\, \textrm{sin}\, \tau}{1 + \textrm{sin}\, \tau \sqrt{1 + \widetilde{r}^2}} \right)\, ,
\end{split}
\end{align}
the NHEK metric becomes
\be
  ds^2 = 2 J \widetilde{\Omega}^2 \left[- (1 + \widetilde{r}^2)\, d\tau^2 + \frac{d\widetilde{r}^2}{1 + \widetilde{r}^2} + d\theta^2 + \widetilde{\Lambda}^2\, (d\varphi + \widetilde{r}\, d\tau)^2 \right]\, .
\ee
This geometry has an $SL(2,\, \mathbb{R}) \times U(1)$ isometry group,  which can be uplifted to a $Vir_L \otimes Vir_R$ symmetry with the central charges $c_L = c_R = 12 J$.

In addition,  all quantum fields on the NHEK background \eqref{eq:NHEK} can be expanded in the modes $e^{- i \omega t + i m \phi}$,  which after taking the near-horizon scaling \eqref{eq:NHscaling} become
\be
  e^{- i \omega t + i m \phi} = e^{- \frac{i}{\lambda} (2 M \omega - m)\, \hat{t} + i m \hat{\phi}} \equiv e^{- i n_R \hat{t} + i n_L \hat{\phi}}\, ,
\ee
from which we can read off the left-moving and the right-moving mode numbers
\be\label{eq:ModeNum}
  n_L \equiv m\, ,\quad n_R \equiv \frac{2 M \omega - m}{\lambda}\, .
\ee
The Boltzmann factor is
\be\label{eq:BoltzmannFactor}
  e^{- \frac{\omega - m \Omega}{T_H}} = e^{- \frac{n_L}{T_L} - \frac{n_R}{T_R}}\, ,
\ee
where $T_H$ is the Hawking temperature, and $T_{L, R}$ are the left-moving and the right-moving Frolov-Thorne temperatures.  By comparing \eqref{eq:ModeNum} and \eqref{eq:BoltzmannFactor},  we obtain
\be
  T_L = \frac{2 M T_H}{1 - 2 M \Omega}\, ,\quad T_R = \frac{2 M T_H}{\lambda}\, .
\ee
The values of $T_{L,  R}$ for the extremal Kerr black holes can be obtained by taking $T_H \to 0$.

Knowing the central charges $c_{L,  R}$ and the Frolov-Thorne temperatures $T_{L,  R}$,  the entropy of a Kerr black hole can be computed using the Cardy formula in 2d CFT:
\be
  S_{BH} = \frac{\pi^2}{3}\, c_L\, R_L + \frac{\pi^2}{3}\, c_R\, T_R\, ,
\ee
where $T_R = 0$ for an extremal Kerr black hole.  The result from this expression exactly matches the one from the Bekenstein-Hawking entropy \eqref{eq:S_BH} on the gravity side.

In the literature,  various generalizations of the original Kerr/CFT correspondence exist for an extremal Kerr black hole.  For instance,  it has been generalized to the Kac-Moody algebra instead of the Virasoro algebra \cite{Compere:2013bya} and to the asymptotically AdS extremal Kerr black holes \cite{Lu:2008jk,  David:2020ems}.

{\flushleft $\bullet$ Near-extremal Kerr/CFT:}

The Kerr/CFT correspondence initially works for extremal Kerr black holes,  which was later generalized to near-extremal Kerr black holes \cite{Bredberg:2009pv,  Hartman:2009nz}.  The absorption cross section of a field scattered off a near-extremal Kerr black hole can be computed using the 2-point function in the corresponding CFT,  and the result exactly matches the one from the gravity approach.

The near-extremal Kerr/CFT correspondence has also been generalized from asymptotically flat black holes to asymptotically AdS black holes \cite{Chen:2010bh},  based on which some physical quantities (Hawking radiation rate,  shear viscosity,  etc.) have been computed \cite{Nian:2020qsk,  David:2020jhp,  Nian:2020bzf}.

{\flushleft $\bullet$ Non-extremal extension of Kerr/CFT:}

For a generic non-extremal Kerr black hole,  the near-horizon conformal symmetry is not manifest as a spacetime symmetry.  Instead,  a conformal symmetry can be revealed in wave equations,  called the hidden conformal symmetry \cite{Castro:2010fd}.  More precisely,  the wave equation of a massless scalar field,  $\frac{1}{\sqrt{-g}}\, \partial_\mu \left(\sqrt{-g}\, g^{\mu\nu}\, \partial_\nu \Phi \right) = 0$,  in the near-horizon region and the low-energy limit can be expressed as
\be
  \mathcal{H}^2\, \Phi = \bar{\mathcal{H}}^2\, \Phi = \ell (\ell + 1) \Phi\, , 
\ee
where $\mathcal{H}^2 = \bar{\mathcal{H}}^2$ are the quadratic Casimirs of two $SL(2,  \mathbb{R})$'s,  indicating the existence of an $SL(2,  \mathbb{R})_L \times SL(2,  \mathbb{R})_R$ conformal symmetry with the central charges $c_L = c_R = 12 J$ and the temperatures $T_{L,  R} = (r_+ \pm r_-) / (4 \pi a)$.  Using these CFT data,  we can apply the Cardy formula in 2d CFT to reproduce the Bekenstein-Hawking entropy \eqref{eq:S_BH} of a non-extremal Kerr black hole.

The hidden conformal symmetry can be understood as the symmetry acting on the soft modes.  By applying the covariant phase space formalism,  one can obtain the same result of the hidden conformal symmetry in a more rigorous way \cite{Haco:2018ske},  including the following CFT data:
\be\label{eq:DefcLcRNonExt}
  c_L = c_R = 12 J\, ,
\ee
\be\label{eq:DefTLTRNonExt}
  T_L = \frac{M^2}{2 \pi J} = \frac{M}{2 \pi a}\, ,\quad T_R = \frac{\sqrt{M^4 - J^2}}{2 \pi J} = \frac{\sqrt{M^2 - a^2}}{2 \pi a}\, ,
\ee
\be\label{eq:DefwLwRNonExt}
  \omega_L = \frac{2 M^3}{J}\, \omega = \frac{2 M^2 \omega}{a}\, ,\quad \omega_R = \frac{2 M^3}{J}\, \omega - m = \frac{2 M^2 \omega}{a} - m\, .
\ee
Consequently,  the entropy of a non-extremal Kerr black hole can be reproduced by the Cardy formula in 2d CFT.

{\flushleft $\bullet$ Soft and hard radiations:}

Now,  let us briefly discuss a black hole's soft and hard radiations.  In a series of papers \cite{Hawking:2016msc,  Hawking:2016sgy},  Hawking,  Perry,  and Strominger have studied extensively the soft hair of a black hole,  i.e.,  some additional conserved charges corresponding to supertranslations and superrotations.  These findings provide new insights into the asymptotic symmetries of black holes and potentially the black hole information paradox.  Further studies show that an appropriate regularization of soft hair can lead to a finite entropy of black holes \cite{Haco:2018ske}.

   \begin{figure}[!htb]
      \begin{center}
        \includegraphics[width=0.37\textwidth]{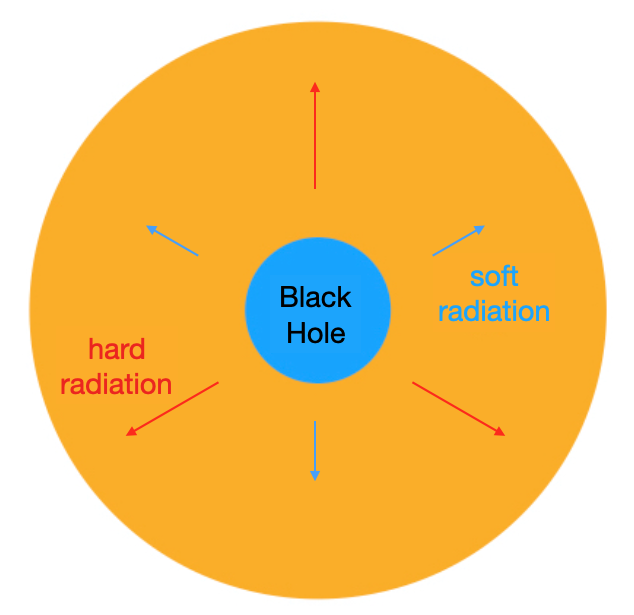}
        \caption{The soft and the hard radiations from a black hole}
        \label{fig:SoftHardRadiations}
      \end{center}
    \end{figure}

However,  soft radiations are insufficient to account for the black hole evaporation.  To carry energy away from a black hole,  the Hawking radiation should include both soft and hard radiations (see Fig.~\ref{fig:SoftHardRadiations}).  They are in a mutualistic relationship.  As we will see in the next section,  the Hawking radiation rate can be computed using a 2d CFT approach,  while the 2d CFT data can be obtained by studying the soft radiations \cite{Haco:2018ske}.  The hard radiations from a black hole follow the radiation rate determined by the soft radiations,  but the hard radiations will bring the black hole to a new state, which has new data on soft radiations.

More precisely,  the soft radiations contain the instantaneous information of the black hole,  including its CFT central charges and Frolov-Thorne temperatures at a fixed time in a CFT$_2$ description,  and they govern the hard radiations via the hidden conformal symmetry or, equivalently, the covariant phase space formalism.  In contrast,  the hard radiations will bring the black hole from the current state to the next state with smaller mass and smaller angular momentum,  i.e.,  they cause the black hole to evaporate.  We must combine soft and hard radiations to have a complete picture of a Kerr black hole evaporation.

\subsection{Hawking Radiation Rate and Conformal Field Theory}\label{sec:CFTinterpretation}

We will see in Sec.~\ref{sec:SemiClassPageCurve} that the temperature of a Kerr black hole increases monotonically during its evaporation process.  Hence,  even for a Kerr black hole initially very close to extremality, it has to be treated as a non-extremal one soon after it starts to emit Hawking quanta and evaporate.  We need to employ the previously described qualitative picture to study the evaporation process of a Kerr black hole,  i.e.,  the soft photons plus the hard photons.

In this subsection,  we will reinforce this picture with some explicit computations from the CFT side.  We will show that the Hawking radiation rate computed via the gravity approach in Sec.~\ref{eq:RateViaGravity} can be reinterpreted in a CFT language,  which is encoded in the information of soft radiations but guides the hard radiation rate \eqref{eq:RateFromGravity} previously derived from the gravity side.

The computation of a black hole's Hawking radiation using the CFT approach was initiated in the work by Emparan and Sachs \cite{EmparanSachs},  where they considered the 3d BTZ black holes.  It is known in the literature that the BTZ black holes are black holes in AdS$_3$ space,  which have dual CFT$_2$ descriptions.  In \cite{EmparanSachs},  the Hawking radiation rate of the BTZ black holes has been computed using the dual CFT$_2$.  More explicitly,  they considered a massless scalar field of frequency $\omega$ and angular momentum quantum number $m$ in a BTZ black hole background.  For simplicity,  $m$ is set to $0$.  Suppose a BTZ black hole has temperatures $(T_L,\, T_R)$ in a CFT$_2$ description.  The CFT approach leads to the Hawking radiation rate of a scalar particle for the BTZ black hole \cite{EmparanSachs}:
\begin{align}\label{eq:EmparanSachsResult}
  \Gamma & = \frac{\omega \pi^2 L^2}{\left[\textrm{exp} \left(\frac{\omega}{2 T_L} \right) - 1 \right]\, \left[\textrm{exp} \left(\frac{\omega}{2 T_R} \right) - 1 \right]}\, ,\nonumber\\
  {} & = \frac{4 \pi^2}{9}\, \frac{\omega\, c_L\, c_R}{\left[\textrm{exp} \left(\frac{\omega}{2 T_L} \right) - 1 \right]\, \left[\textrm{exp} \left(\frac{\omega}{2 T_R} \right) - 1 \right]}\, ,
\end{align}
where $L$ is the AdS$_3$ radius,  and we have used the relation $c = 3 L / (2 G_N)$ in AdS$_3$/CFT$_2$ (with $G_N = 1$).  The second line of \eqref{eq:EmparanSachsResult} manifestly depends on the CFT$_2$ data and is essentially the particle number flux of a perfect blackbody radiation in (2+1) dimensions.

As reviewed in the previous subsection,  a Kerr black hole always has some near-horizon conformal symmetry,  regardless of whether it is extremal.  Hence,  we can use the CFT approach to compute the Hawking radiation rate of non-extremal Kerr black holes,  similar to the treatment of the BTZ black hole in \cite{EmparanSachs}.  In the rest of this subsection,  we will focus on the patch of Region $I$ introduced in Fig.~\ref{fig:KerrBHPenroseDiagram3Regions} and consider the Hawking radiation rate from the black hole side measured at the future infinity of Region $I$.  We will see that the results in Sec.~\ref{eq:RateViaGravity} using the gravity approach have CFT interpretations.  To simplify the discussion,  we only consider a scalar field ($s = 0$) in this subsection.

\subsubsection{The Near-Horiozn Region}

Since a non-extremal Kerr black hole has a hidden conformal symmetry,  we can apply the CFT approach similar to \cite{EmparanSachs} and compute the Hawking radiation rate.  This calculation of Hawking radiation for a non-extremal Kerr black hole is essentially the same as the computation of gravitational waves in \cite{Nian:2023dng} but with different boundary conditions.

Before discussing the CFT approach,  let us take a closer look at the near-horizon solution from the gravity side.  The near-horizon limit $\omega r \ll 1$ has been considered in \cite{Castro:2010fd,  Nian:2023dng},  which is related but slightly different from the near-horizon limit $k x \ll l + 1$ discussed in Sec.~\ref{eq:RateViaGravity}.  For the former one,  the radial part of a massless scalar wave equation with purely ingoing boundary conditions at the horizon has been obtained in \cite{Castro:2010fd,  Nian:2023dng}:
\begin{align}\label{eq:NearSol in HiddenConfSymm}
  R (r) & = \left(\frac{r - r_+}{r - r_-} \right)^{-i\, \frac{2 M r_+ \omega - a m}{r_+ - r_-}}\, \left(1 - \frac{r - r_+}{r - r_-} \right)^{1 + l}\nonumber\\
  {} & \quad \cdot \phantom{|}_2 F_1 \left(1 + l - i\, \frac{4 M^2 \omega - 2 a m}{r_+ - r_-},\, 1 + l - i\, 2 M \omega;\, 1 - i\, \frac{4 M r_+ \omega - 2 a m}{r_+ - r_-};\, \frac{r - r_+}{r - r_-} \right)\, ,
\end{align}
which has the asymptotic behavior for $r \gg M$ \cite{Castro:2010fd,  Nian:2023dng}:
\be\label{eq:AsymptoticBehavior}
  R \to A\, (r_+ - r_-)^{1 + l}\, r^{- 1 - l} + B\, (r_+ - r_-)^{-l}\, r^l\, ,
\ee
with
\begin{align}
\begin{split}\label{eq:DefAandB}
  A & = \frac{\Gamma\left(1 - i\, \frac{4 M r_+ \omega - 2 a m}{r_+ - r_-} \right)\cdot \Gamma\left(- 1 - 2 l \right)}{\Gamma\left(- l - i\, \frac{4 M^2 \omega - 2 a m}{r_+ - r_-} \right)\cdot \Gamma\left(- l - 2 i M \omega \right)}\, ,\\
  B & = \frac{\Gamma\left(1 - i\, \frac{4 M r_+ \omega - 2 a m}{r_+ - r_-} \right)\cdot \Gamma\left(1 + 2 l \right)}{\Gamma\left(1 + l - i\, \frac{4 M^2 \omega - 2 a m}{r_+ - r_-} \right)\cdot \Gamma\left(1 + l - 2 i M \omega \right)}\, .
\end{split}
\end{align}
We have seen in Sec.~\ref{eq:RateViaGravity} that the radial wave equation for the slightly different near-horizon limit $k x \ll l + 1$ has the solution \eqref{eq:NearSol},  which for the spin $s=0$ becomes
\begin{align}\label{eq:NearSol by Page}
  R(r) & = \left(\frac{r - r_+}{r - r_-} \right)^{i Q}\phantom{|}_2 F_1 \left(-l,\, 1 + l;\, 1 + 2 i Q;\, \frac{r - r_+}{r_- - r_+} \right)\nonumber\\
  {} & = \left(\frac{r - r_+}{r - r_-} \right)^{i Q}\, \left(1 - \frac{r - r_+}{r - r_-} \right)^{1 + l}\phantom{|}_2 F_1 \left(1 + l + 2 i Q,\, 1+l;\, 1 + 2 i Q;\, \frac{r - r_+}{r - r_-} \right)\, ,
\end{align}
where we have used the following identity:
\be
  \phantom{|}_2 F_1 \left(a,\, b;\, c;\, z \right) = (1 - z)^{-b}\phantom{|}_2 F_1 \left(c - a,\, b;\, c;\, \frac{z}{z - 1} \right)\, .
\ee
Taking into account the definitions \eqref{eq:Definitions},  it is easy to check that the two solutions \eqref{eq:NearSol in HiddenConfSymm} and \eqref{eq:NearSol by Page} coincide at the leading order in the low-energy limit $M \omega \ll 1$.  Hence,  the solution \eqref{eq:NearSol by Page} has the same asymptotic behavior \eqref{eq:AsymptoticBehavior} at the leading order.  The two terms in \eqref{eq:AsymptoticBehavior} correspond to the Neumann and the Dirichlet boundary conditions,  respectively.

The full solution to the scalar wave equation in the near-horizon region is
\be
  \Psi = \sum_{l,\, m} e^{- i \omega t + i m \phi}\, R_{lm} (r)\, S_l (\theta)\, ,
\ee
where the radial part $R_{lm}$ is given by \eqref{eq:NearSol in HiddenConfSymm}.  Based on this solution,  we can compute the particle flux at the horizon $r_+$ defined by
\be
  \mathcal{F} \equiv \int \sqrt{-g}\, J^r\, d\theta\, d\phi\, ,\quad \text{with } J^\mu \equiv \frac{i}{8 \pi} \left(\Psi^* \nabla^\mu \Psi - \Psi \nabla^\mu \Psi^* \right)\, ,
\ee
and the result in this case is
\be\label{eq:FlmNearHorizon}
  \mathcal{F} = \sum_{l,\, m} \mathcal{F}_{lm}\, ,\quad \text{with } \mathcal{F}_{lm} = \frac{1}{2} (2 M r_+ \omega - a m)\, .
\ee

Now,  let us turn to the CFT treatment.  Suppose the dual CFT$_2$ is deformed by some operators $\mathcal{O}_l$.  The corresponding action of the deformed theory is
\be\label{eq:CFTdeformation}
  S = S_{CFT} + \sum_{l,\, m} \int dt^+\, dt^-\, J_{l m}\, e^{i \omega_L t^+ - i \omega_R t^-}\, \mathcal{O}_l (t^+,\, t^-)\, .
\ee
The Hawking radiations can be viewed as transitions induced by these deformation operators.  Similar to \cite{Maldacena:1997ih,  Gubser:1997cm,  EmparanSachs},  the radiation rate at the horizon can be computed using Fermi's golden rule,  and the result is
\be\label{eq:FermiGoldenRule}
 \mathcal{R} = 2 \pi \sum_{l,\, m} \Big|J_{l m} \Big|^2\, \int dt^+\, dt^-\, e^{i \omega_L t^+ - i \omega_R t^-}\, \langle \mathcal{O}_{(l,  l)}^\dagger (t^+,\, t^-)\, \mathcal{O}_{(l,  l)} (0,\, 0) \rangle_{T_L,\, T_R}\, ,
\ee
where $\langle \mathcal{O}_{(h_L,  h_R)}^\dagger (t^+,\, t^-)\, \mathcal{O}_{(h_L,  h_R)} (0,\, 0) \rangle_{T_L,\, T_R}$ is the 2-point correlation function of the CFT$_2$ operator $\mathcal{O}$ with conformal weights $(h_L,  h_R)$ at finite temperatures $T_{L,\, R}$:
\be
  \langle \mathcal{O}_{(h_L,  h_R)}^\dagger (t^+,\, t^-)\, \mathcal{O}_{(h_L,  h_R)} (0,\, 0) \rangle_{T_L,\, T_R} = \frac{C_\mathcal{O}^2}{i^{2 h_L + 2 h_R}} \left(\frac{\pi T_L}{\textrm{sinh} \left(\pi T_L t^+ \right)} \right)^{2 h_L} \left(\frac{\pi T_R}{\textrm{sin} \left(\pi T_R t^- \right)} \right)^{2 h_R}\, ,
\ee
and the source $J_{l m}$ can be obtained by extending the solution in the near-horizon region ($\omega r \ll 1$) to the asymptotic region ($r \gg M$),  which for the near-horizon solution \eqref{eq:AsymptoticBehavior} with the Dirichlet boundary condition is $J_{l m} = B$ given by \eqref{eq:DefAandB}.

Using the following identity
\be
  \int dx\, e^{\pm i \omega x}\, \left(\frac{\pi T}{\textrm{sinh} \left(\pi T x \right)} \right)^{2 \Delta} = (-1)^\Delta\, \frac{(2 \pi T)^{2 \Delta - 1}}{\Gamma (2 \Delta)}\, e^{\pm \omega / 2 T}\, \bigg|\Gamma\left(\Delta + i \frac{\omega}{2 \pi T} \right) \bigg|^2\, ,
\ee
with a positive integer $2 \Delta$,  we can evaluate the integrals in \eqref{eq:FermiGoldenRule} and obtain
\be
  \mathcal{R} = \sum_{l,\, m} \mathcal{R}_{lm}\, ,
\ee
with
\begin{equation}
\begin{aligned}
  \mathcal{R}_{lm} & = 4\pi^2C_\mathcal{O}^2\left(\frac{M}{a}\right)^{2l-1}\left(\frac{\sqrt{M^2-a^2}}{a}\right)^{2l-1}\cdot \frac{2Mr_+\omega - am}{r_+ - r_-} \cdot \frac{e^{\pi\frac{4M\omega r_+-2am}{r_+-r_-}}}{\sinh\pi\frac{4Mr_+\omega-2am}{r_+-r_-}} \nonumber\\
{} & \quad\cdot \frac{4l^2}{\left(l^2+\left(\frac{4M^2\omega-2am}{r_+-r_-}\right)^2\right)\, \left(l^2+(2M\omega)^2\right)}\\
{} & \approx \frac{16\pi^2}{al^2}C_\mathcal{O}^2\left(\frac{M}{a}\right)^{2l-1}\left(\frac{\sqrt{M^2-a^2}}{a}\right)^{2l-2}(2Mr_+\omega-am)\cdot \frac{1}{1 - e^{- 2\pi\frac{4Mr_+\omega-2am}{r_+-r_-}}}\, ,
\end{aligned}
\end{equation}
where in the last expression we only keep the terms at the leading order in $(\omega,  m)$,  and $C_\mathcal{O}$ is a normalization constant for the operator $\mathcal{O}$.  By choosing an appropriate constant $C_\mathcal{O}$ independent of $(\omega,  m)$:
\be
  C_\mathcal{O} = \frac{\sqrt{a}\, l}{4 \sqrt{2} \pi} \left(\frac{a}{M} \right)^{l - \frac{1}{2}} \left(\frac{a}{\sqrt{M^2 - a^2}} \right)^{l - 1}\, ,
\ee
we obtain $\mathcal{R}_{lm}$ at the leading order in $(\omega,  m)$:
\be\label{eq:RlmNearHorizon}
  \mathcal{R}_{lm} = \frac{1}{2} (2 M r_+ \omega - a m)\cdot \frac{1}{1 - e^{- \beta_L \omega_L - \beta_R \omega_R}}\, ,
\ee
where $\omega_{L,  R}$ are defined in \eqref{eq:DefwLwRNonExt},  and $\beta_{L,  R} \equiv 1 / T_{L,  R}$ with $T_{L,  R}$ defined in \eqref{eq:DefTLTRNonExt}.  Interestingly,  this expression has the same form as the particle number radiation rate of a (2+1)d blackbody,  reflecting the nature of a CFT$_2$ result naturally related to (2+1)d gravity.

By comparing \eqref{eq:FlmNearHorizon} and \eqref{eq:RlmNearHorizon},  we obtain a relation between the gravity and the CFT results:
\be\label{eq:MatchAtNearHorizon}
  \mathcal{F}_{lm} = \mathcal{R}_{lm}\, \left(1 - e^{- \beta_L \omega_L - \beta_R \omega_R} \right)\, ,
\ee
which can be interpreted as follows.  The transition rate $\mathcal{R}_{lm}$ obtained from the CFT approach contains the stimulated emissions.  To compare with the flux $\mathcal{F}_{lm}$,  the contributions from the stimulated emissions should be subtracted from $\mathcal{R}_{lm}$.  Taking this factor into account,  we find an exact match between the gravity and the CFT results.  Similar discussions can be found in Refs.~\cite{Gubser:1997cm,  Nian:2023dng}.

\subsubsection{The Asymptotic Future Infinity}

We have seen that the radiation rate in the near-horizon region matches the flux from the gravity approach.  However,  the standard definition of the Hawking radiation rate is at the asymptotic future infinity,  and its computation from the gravity side has been reviewed in Sec.~\ref{eq:RateViaGravity}.  A completely independent CFT computation of the Hawking radiation rate at the infinity is not available yet.  Nevertheless,  we can provide a CFT interpretation for the gravity result.

The main idea is to extrapolate the asymptotic solution to the near-horizon region and find its CFT interpretation there,  similar to the one mentioned before. The crucial step is to specify the boundary conditions carefully.  More precisely,  we consider a CFT deformed by some operators as in \eqref{eq:CFTdeformation},  and then compute the radiation rate induced by the deformation operators using Fermi's golden rule as in \eqref{eq:FermiGoldenRule}.  The difference is how to determine the source term $J_{lm}$.  Suppose that two independent solutions to the radial wave equation in the near region are $R^{(1),  (2)}$,  while the two in the far region are $\widetilde{R}^{(1),  (2)}$.  We will apply the following procedure,  different from the one used in \cite{Page1976} (also reviewed in Sec.~\ref{eq:RateViaGravity}):
\begin{enumerate}
  \item Consider the $(r \to \infty)$ asymptotic behaviors of $\widetilde{R}^{(1),  (2)}$,  i.e.,  $\widetilde{R}^{(1),  (2)}_{r \to \infty}$.  For a given asymptotic boundary condition,  we can fix the coefficients $(\widetilde{P},  \widetilde{Q})$ in the linear combination
  \be
    \widetilde{R}^{(3)}_{r \to \infty} = \widetilde{P}\, \widetilde{R}^{(1)}_{r \to \infty} + \widetilde{Q}\, \widetilde{R}^{(2)}_{r \to \infty}\, ;
  \ee
  
  \item Glue the near solution $R^{(3)} = A\, R^{(1)} + B\, R^{(2)}$ and the far solution $\widetilde{R}^{(3)} = \widetilde{P}\, \widetilde{R}^{(1)} + \widetilde{Q}\, \widetilde{R}^{(2)}$ in the overlapping region to fix the coefficients $(A,  B)$;
  
  \item The source term $J_{lm}$ can be read off from the $(r \to \infty)$ asymptotic behavior of the near solution $R^{(3)}$.
\end{enumerate}
In contrast,  the approach of \cite{Page1976} (reviewed in  Sec.~\ref{eq:RateViaGravity}) first fixes the near solution $R^{(3)}$ with a certain boundary condition,  and then eventually determines the $(r \to \infty)$ asymptotic behavior of the far solution $\widetilde{R}^{(3)}$.

The starting point is similar to the treatment of gravitational waves in \cite{Nian:2023dng}.  For the far region ($r \gg M$),  the radial wave equation has two independent solutions:
\be
  \widetilde{R}^{(1)} = \frac{1}{\sqrt{\omega r}}\, J_{l + \frac{1}{2}} (\omega r)\, ,\quad \widetilde{R}^{(2)} = \frac{1}{\sqrt{\omega r}}\, J_{- l - \frac{1}{2}} (\omega r)\, ,
\ee
which have the asymptotic behaviors
\begin{align}
  \widetilde{R}^{(1)} & = \left\{
  \begin{aligned}
    & \omega^l r^l\, \frac{2^{- l - \frac{1}{2}}}{\Gamma (l + \frac{3}{2})}\, , \quad\quad\qquad\,\, \textrm{for } r \to 0\, ; \\
    & \sqrt{\frac{2}{\pi}}\, \frac{1}{\omega r}\, \textrm{sin} \left(\omega r - \frac{\pi l}{2}\right)\, , \quad \textrm{for } r \to \infty\, ,
  \end{aligned} \right.\\
  \widetilde{R}^{(2)} & = \left\{
  \begin{aligned}
    & \omega^{-l-1} r^{-l-1}\, \frac{2^{l + \frac{1}{2}}}{\Gamma (- l + \frac{1}{2})}\, , \,\, \textrm{for } r \to 0\, ; \\
    & \sqrt{\frac{2}{\pi}}\, \frac{1}{\omega r}\, \textrm{sin} \left(\omega r + \frac{\pi l}{2}\right)\, , \quad \textrm{for } r \to \infty\, .
  \end{aligned} \right.
\end{align}
The general solution can be a linear combination:
\be
  \widetilde{R}^{(3)} = \widetilde{P}\cdot \widetilde{R}^{(1)} + \widetilde{Q}\cdot \widetilde{R}^{(2)}\, ,
\ee
where $\widetilde{P}$ and $\widetilde{Q}$ are two constant coefficients.  For the solution to be purely outgoing at infinity,  there should be a condition
\be
  \widetilde{P} = i e^{- i \pi l} \widetilde{Q}\, .
\ee
Under this condition,  the general solution at the asymptotic infinity becomes
\be
  \widetilde{R}^{(3)} \to \widetilde{R}^{(3)}_{r \to \infty} = \widetilde{Q}\, \sqrt{\frac{2}{\pi}}\, e^{-\frac{i}{2} (l \pi - 2 \omega r)}\, \frac{\textrm{cos} (l \pi)}{\omega r}\, .
\ee
Assuming that $l$ is an integer,  we can show that the particle number flux at the infinity is $\propto \widetilde{Q}^2 / \omega$,  which is expected to be proportional to the Hawking radiation probability $\Gamma$ obtained from the gravity side in \eqref{eq:Gamma}.  Hence,  $\widetilde{Q} = \widetilde{C} \sqrt{\Gamma\, \omega}$,  where $\widetilde{C}$ is an $(\omega,  m)$-independent constant.  The far-region solution obeying the outgoing boundary condition and reproducing the correct Hawking radiation rate at infinity is
\be
  \widetilde{R}^{(3)} = \widetilde{Q} \left(i e^{- i \pi l}\, \widetilde{R}^{(1)} + \widetilde{R}^{(2)} \right)\, ,
\ee
It has the following leading order behavior at $r \to 0$:
\be
  \widetilde{R}^{(3)} \to \widetilde{R}^{(3)}_{r \to 0} \approx \widetilde{Q}\, \omega^{-l-1} r^{-l-1}\, \frac{2^{l + \frac{1}{2}}}{\Gamma (- l + \frac{1}{2})}\, .
\ee

This $(r \to 0)$ asymptotics of the far-region solution should be glued with the $(r \to \infty)$ asymptotics of the near-region solution,  which fixes the near-region solution to be
\be\label{eq:R(3)}
  R^{(3)} = C_0 \left(D\cdot R^{(1)} - B\cdot R^{(2)} \right)\quad \text{with }\,\, C_0 \equiv\frac{\widetilde{Q}\, \omega^{-l-1}}{(A D - B C) (r_+ - r_-)^{l + 1}} \frac{2^{l + \frac{1}{2}}}{\Gamma (- l + \frac{1}{2})}\, ,
\ee
where $R^{(1),  (2)}$ are defined as \cite{Nian:2023dng}:
\begin{align}
  R^{(1)} (r) & = \left(\frac{r - r_+}{r - r_-} \right)^{-i\, \frac{2 M r_+ \omega - a m}{r_+ - r_-}}\, \left(1 - \frac{r - r_+}{r - r_-} \right)^{1 + l}\nonumber\\
  {} & \quad \cdot \phantom{|}_2 F_1 \left(1 + l - i\, \frac{4 M^2 \omega - 2 a m}{r_+ - r_-},\, 1 + l - i\, 2 M \omega;\, 1 - i\, \frac{4 M r_+ \omega - 2 a m}{r_+ - r_-};\, \frac{r - r_+}{r - r_-} \right)\, ,\\
  R^{(2)} (r) & = \left(\frac{r - r_+}{r - r_-} \right)^{i\, \frac{2 M r_+ \omega - a m}{r_+ - r_-}}\, \left(1 - \frac{r - r_+}{r - r_-} \right)^{1 + l}\nonumber\\
  {} & \quad \cdot \phantom{|}_2 F_1 \left(1 + l + i\, \frac{4 M^2 \omega - 2 a m}{r_+ - r_-},\, 1 + l + i\, 2 M \omega;\, 1 + i\, \frac{4 M r_+ \omega - 2 a m}{r_+ - r_-};\, \frac{r - r_+}{r - r_-} \right)\, .
\end{align}
The coefficients $(A,  B)$ have been defined in \eqref{eq:DefAandB},  while $(C,  D)$ are defined as follows:
\begin{align}
\begin{split}\label{eq:DefCandD}
  C & = \frac{\Gamma\left(1 + i\, \frac{4 M r_+ \omega - 2 a m}{r_+ - r_-} \right)\cdot \Gamma\left(- 1 - 2 l \right)}{\Gamma\left(- l + i\, \frac{4 M^2 \omega - 2 a m}{r_+ - r_-} \right)\cdot \Gamma\left(- l + 2 i M \omega \right)}\, ,\\
  D & = \frac{\Gamma\left(1 + i\, \frac{4 M r_+ \omega - 2 a m}{r_+ - r_-} \right)\cdot \Gamma\left(1 + 2 l \right)}{\Gamma\left(1 + l + i\, \frac{4 M^2 \omega - 2 a m}{r_+ - r_-} \right)\cdot \Gamma\left(1 + l + 2 i M \omega \right)}\, .
\end{split}
\end{align}
The two independent solutions $R^{(1)}$ and $R^{(2)}$ have different boundary conditions at the horizon $r_+$: $R^{(1)}$ is purely ingoing,  while $R^{(2)}$ is purely outgoing.  To find the source term with appropriate boundary conditions,  we should consider the $R^{(1)}$-part contained in $R^{(3)}$,  i.e.,  $C_0\, D\, R^{(1)}$,  and extend it to $r \to \infty$.  The asymptotic behavior of the solution $R^{(1)}$ at $r \to \infty$ splits into two branches satisfying the Neumann and the Dirichlet boundary conditions,  respectively:
\be
  R^{(1)} \to A\, (r_+ - r_-)^{l + 1}\, r^{- l - 1} + B\, (r_+ - r_-)^{-l}\, r^l\, ,\quad \text{for } r \to \infty\, .
\ee
The source term $J_{l m}$ we are looking for is given by the asymptotic solution at $r \to \infty$ with the Dirichlet boundary condition:
\be\label{eq:JlmNew}
  J_{l m} = C_0\, D\, B\, .
\ee

The full solution near the horizon $r_+$ is $\Psi = \sum_{l,\, m} e^{- i \omega t + i m \phi}\, R_{lm} (r)\, S_l (\theta)$ with $R_{lm}$ given by \eqref{eq:R(3)}.  Using the same method in the previous subsection,  we can evaluate the particle flux at the horizon $r_+$:
\be\label{eq:FlmNew}
  \mathcal{F} = \sum_{l,\, m} \mathcal{F}_{lm}\, ,\quad \text{with } \mathcal{F}_{lm} = \frac{1}{2} (2 M r_+ \omega - a m)\cdot |C_0\, D|^2\, .
\ee
Similarly,  applying the source term \eqref{eq:JlmNew} to Fermi's golden rule \eqref{eq:FermiGoldenRule},  we can evaluate the near-region radiation rate at the leading order in $(\omega,\, m)$:
\be\label{eq:RlmNew}
  \mathcal{R} = \sum_{l,\, m} \mathcal{R}_{lm}\, ,\quad \text{with } \mathcal{R}_{lm} = \frac{1}{2} (2 M r_+ \omega - a m)\cdot \frac{|C_0\, D|^2}{1 - e^{- \beta_L \omega_L - \beta_R \omega_R}}\, .
\ee
Comparing the flux at the horizon $r_+$ \eqref{eq:FlmNew} with the near-region radiation rate \eqref{eq:RlmNew},  we again find that
\be
  \mathcal{F}_{lm} = \mathcal{R}_{lm} (1 - e^{- \beta_L \omega_L - \beta_R \omega_R})\, .
\ee
Importantly,  both $\mathcal{F}_{lm}$ and $\mathcal{R}_{lm}$ are proportional to the Hawking radiation probability $\Gamma$ at infinity obtained in \eqref{eq:Gamma} from the gravity side.  In particular,  the near-region radiation rate $\mathcal{R}_{lm}$ even has the Bose-Einstein distribution factor.  Hence,  it is proportional to the particle rate of the Hawking radiation obtained in \eqref{eq:RateFromGravity}.

In summary,  the correspondence between the flux at the horizon $r_+$ and the radiation rate holds only in the near-horizon region.  By gluing the gravity solutions in different regions using their asymptotic behaviors,  we find that the Hawking radiation probability $\Gamma$ in asymptotic infinity is encoded in the near-region solutions.  Although the CFT approach is not an independent method in this case,  it can naturally provide the Bose-Einstein distribution factor in the radiation rate.  In contrast,  this factor was introduced by hand in the gravity approach.

\subsection{Entanglement Entropy and Information Paradox}

In this section,  we have seen the computation of the Hawking radiation from the gravity side and its CFT counterpart.  Although the CFT approach is not completely independent,  rewriting the result in the CFT language still provides some insight into the problem.  For instance,  the factor of the Bose-Einstein statistics shows up naturally from the CFT approach,  while it has to be introduced by hand in the gravity approach.

As we have reviewed in Sec.~\ref{eq:KerrCFTReview},  whether extremal or not,  a Kerr black hole possesses a near-horizon conformal symmetry.  In particular,  for the non-extremal Kerr black holes,  the hidden conformal symmetry can be obtained by regularizing the infinitely many soft hairs \cite{Haco:2018ske}.  As discussed before,  the soft hair is similar to the quantum numbers that can distinguish degenerate states in a hydrogen atom,  while the soft and the hard radiations can be schematically understood as the photon emissions between the degenerate and the non-degenerate states in a hydrogen atom.  Hence,  the full picture of the radiation-induced black hole evaporation is as follows:
\begin{enumerate}
  \item[1)] The soft radiations do not cause the black hole to evaporate,  but from them, we can read off the current status of the black hole,  for instance,  the CFT data $(c_L,\, c_R)$ and $(T_L,\, T_R)$;
  
  \item[2)] The hard radiations cause the black hole to evaporation.  More precisely,  they take a finite amount of energy and angular momentum from the Kerr black hole.

  \item[3)] The soft radiations govern the hard radiations at this moment,  i.e.,  the CFT data from the soft radiations determine the Hawking radiation rate of the hard radiations.
  
  \item[4)] After emitting some hard radiations,  the black hole arrives at a new state with new soft radiations.
\end{enumerate}
In summary,  the soft and the hard radiations dynamically affect each other: The soft radiations provide the current information of the black hole,  which controls the hard radiation at this moment,  while the hard radiations bring the black hole from the current state to a new state.

In \cite{Nian:2019buz},  we have solved the differential equation system \eqref{eq:ODE} numerically and can describe the whole evaporation process of a Kerr black hole,  i.e.,  for given initial conditions $(M(t),\,  J(t))$ are known functions of the time $t$.

\begin{figure}[!htb]
      \begin{center}
        \includegraphics[width=0.71\textwidth]{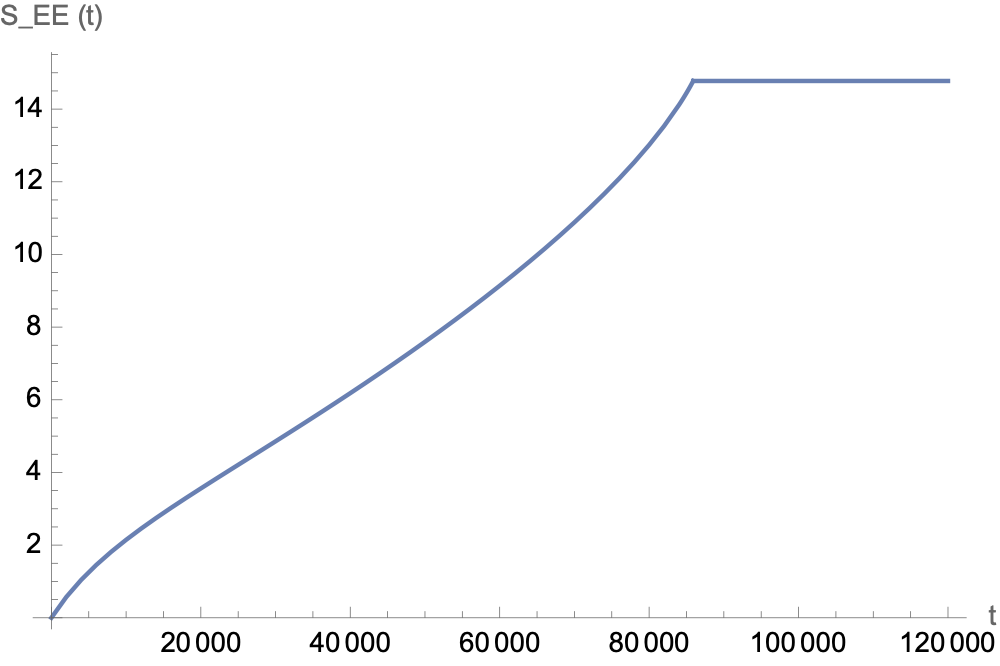}
        \caption{The time evolution of the entanglement entropy between a Kerr black hole (with $M(0) = 1$ and $J(0) = 1/2$) and its emitted Hawking quanta in Hawking's approach}
        \label{fig:SEEfromBH}
      \end{center}
    \end{figure}

From the gravity computation \cite{Page1976},  the particle number rate observed at infinity is
\be\label{eq:dNdt}
  \frac{dN}{dt} = \frac{1}{2 \pi} \sum_{slmp} \int d\omega\, \langle N_{swlmp} \rangle = \frac{1}{2 \pi} \sum_{slmp} \int d\omega\, \Gamma_{s \omega l m p}\, \frac{1}{\textrm{exp} \left[\frac{1}{T_H} (\omega - m \Omega) \right] \mp 1}\, ,
\ee
where $\langle N_{s \omega l m p} \rangle$ and $\Gamma_{s \omega l m p}$ are given in \eqref{eq:RateFromGravity} and \eqref{eq:Gamma},  respectively.  In the following of this paper,  we focus on the photon as the emission particle,  for which $l = s = 1$ has the dominant contribution.  Consequently,  \eqref{eq:dNdt} in this case becomes
\be\label{eq:dNdtPhoton}
  \frac{dN}{dt} = \frac{1}{2 \pi} \sum_{mp} \int d\omega\, \langle N_{1w1mp} \rangle = \frac{1}{2 \pi} \sum_{mp} \int d\omega\, \Gamma_{1 \omega 1 m p}\, \frac{1}{\textrm{exp} \left[\frac{1}{T_H} (\omega - m \Omega) \right] - 1}\, ,
\ee
For one observed photon at infinity,  we assume it is one particle in a maximally entangled pair of Hawking quanta.  The other particle in the entangled pair falls into the black hole,  which creates one-bit entanglement entropy between the black hole and the Hawking quanta.  Hence,  the particle number rate at infinity \eqref{eq:dNdtPhoton} is also proportional to the increasing rate of the entanglement entropy measured at infinity.  By integrating \eqref{eq:dNdtPhoton},  we obtain the time evolution of the entanglement entropy between the black hole and the Hawking quanta,  $S_{EE} (t) = N(t)\, \textrm{log} (2)$.  By choosing an appropriate unit,  we drop the factor $\textrm{log} (2)$ for simplicity in the rest of this paper.  Using this approach,  we numerically compute $S_{EE}^{(BH)} (t)$,  which is shown in Fig.~\ref{fig:SEEfromBH},  where the superscript ``BH'' indicates that the entanglement is induced from the black hole side.  Later,  we will introduce the entanglement loss caused from the white hole side.  Some ideas similar to our approach have been discussed in the literature \cite{Mathur:2009hf,  Han:2023wxg}.

\section{Page Curve of Kerr Black Hole}\label{sec:PageCurve}

In contrast to the monotonically increasing result from Hawking's approach,  Page argued the entanglement entropy between a Kerr black hole and its Hawking radiation quanta during the black hole evaporation would first increase from zero and reach a maximum at a particular time; after that, it decreases to zero again.  This time evolution curve of entanglement entropy is the so-called Page curve.  In this section,  we first review Page's original computation for a Schwarzschild black hole and its generalization to a Kerr black hole.  The central part of this section is to provide a microscopic calculation of the Page curve for a Kerr black hole,  which is also the paper's main result.

\subsection{Semi-Classical Computation of Page Curve}\label{sec:SemiClassPageCurve}

Before discussing the semi-classical computation of the Page curve,  let us first review the evolution of the Bekenstein-Hawking entropy of a Schwarzschild black hole during the evaporation.

Because the Hawking radiation of a (3+1)d Schwarzschild black hole is a greybody radiation,  we obtain the power of the Hawking radiation through the horizon by applying the Stefan-Boltzmann law:
\be
  L = \frac{\Gamma \gamma}{15360\, \pi M^2}\, ,
\ee
where $\Gamma$ and $\gamma$ denote the greybody factor and the total number of massless degrees of freedom,  respectively.  Consequently,  the time evolution of the black hole mass is
\be
  \frac{dM}{dt} = - L\, ,
\ee
which leads to the solution
\be\label{eq:Schwarzschild M(t)}
  M(t) = M_0 \left(1 - \frac{t}{t_L} \right)^\frac{1}{3}\, ,
\ee
where $t_L \equiv 5120\, \pi M_0^3 / \Gamma \gamma$ with $M_0$ denoting the initial mass of the black hole.  Hence,  the time evolution of the black hole entropy is
\be
  S_{BH} (t) = 4 \pi M^2 (t) = 4 \pi M_0^2 \left(1 - \frac{t}{t_L} \right)^\frac{2}{3}\, .
\ee

\begin{figure}[!htb]
      \begin{center}
        \includegraphics[width=0.67\textwidth]{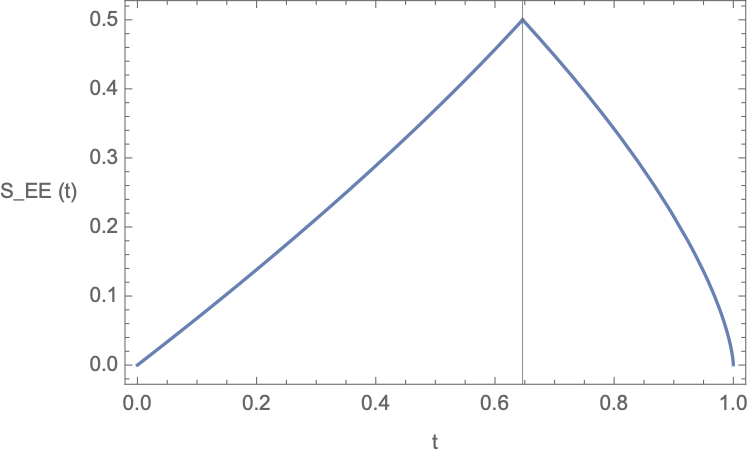}
        \caption{The semi-classical result of the Page curve for a Schwarzschild black hole with $\beta = 1$ and $M_0 = 1 / \sqrt{4 \pi}$}
        \label{fig:SemiClassicalPageCurveSchwarzschild}
      \end{center}
    \end{figure}

As argued in \cite{Page:1993wv,  Page:2013dx},  the entanglement entropy of the Hawking radiation, $S_{EE} (t)$,  should be close to $S_{rad} (t)$ for $t < t_*$ and close to $S_{BH} (t)$ for $t > t_*$.  Therefore, $S_{EE} (t)$ can be approximated as:
\begin{align}\label{eq:ClassicalPageCurve}
  S_{EE} (t) & \approx S_{rad}\cdot \Theta (t_* - t) + S_{BH}\cdot \Theta (t - t_*) \nonumber\\
  {} & = 4 \pi \beta M_0^2 \left[1 - \left(1 - \frac{t}{t_L} \right)^\frac{2}{3} \right]\, \theta (t_* - t) + 4 \pi M_0^2 \left(1 - \frac{t}{t_L} \right)^\frac{2}{3}\, \theta (t - t_*)\, ,
\end{align}
with $\beta \equiv \frac{dS_{rad} / dt}{- dS_{BH} / dt}$ and $t_* \equiv t_L \left[1 - \left(\frac{\beta}{1 + \beta} \right)^{3/2} \right]$.  The expression \eqref{eq:ClassicalPageCurve} corresponds to the Page curve for a Schwarzschild black hole (see Fig.~\ref{fig:SemiClassicalPageCurveSchwarzschild}) with the Page time $t_*$.

\begin{figure}[!htb]
      \begin{center}
        \includegraphics[width=0.67\textwidth]{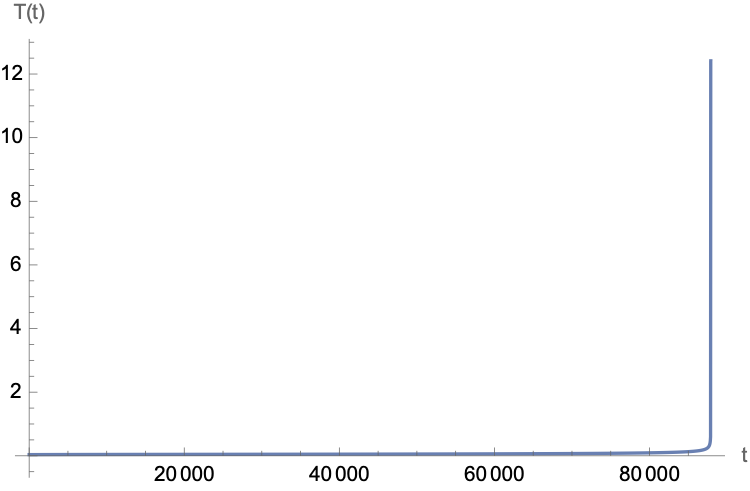}
        \caption{The evolution of the Hawking temperature $T_H (t)$ for a Kerr black hole with the initial conditions $M(0)=1$,  $J(0)=1/2$}
        \label{fig:T(t)}
      \end{center}
    \end{figure}

\begin{figure}[!htb]
      \begin{center}
        \includegraphics[width=0.67\textwidth]{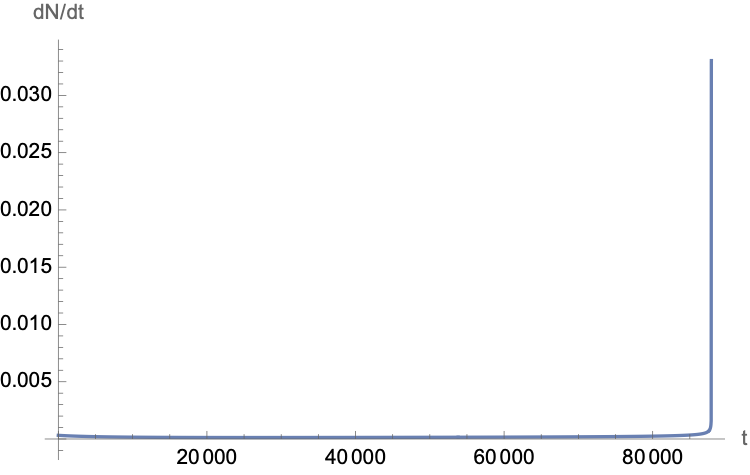}
        \caption{The evolution of the Hawking radiation rate $\frac{dN}{dt} (t)$ for a Kerr black hole with the initial conditions $M(0)=1$,  $J(0)=1/2$}
        \label{fig:dNdt}
      \end{center}
    \end{figure}

\newpage
The semi-classical computation of the Page curve has been generalized to the Kerr black hole in \cite{Nian:2019buz}.  The main idea is first to solve the coupled ordinary differential equation system \eqref{eq:ODE} to find $M(t)$ and $J(t)$.  Knowing $M(t)$ and $J(t)$ as functions of time,  we can compute all the thermodynamic quantities,  for instance,  the black hole temperature $T_H$ and the particle radiation rate $dN/dt$ as functions of time (see Figs.~\ref{fig:T(t)} and \ref{fig:dNdt}). 

\begin{figure}[!htb]
      \begin{center}
        \includegraphics[width=0.67\textwidth]{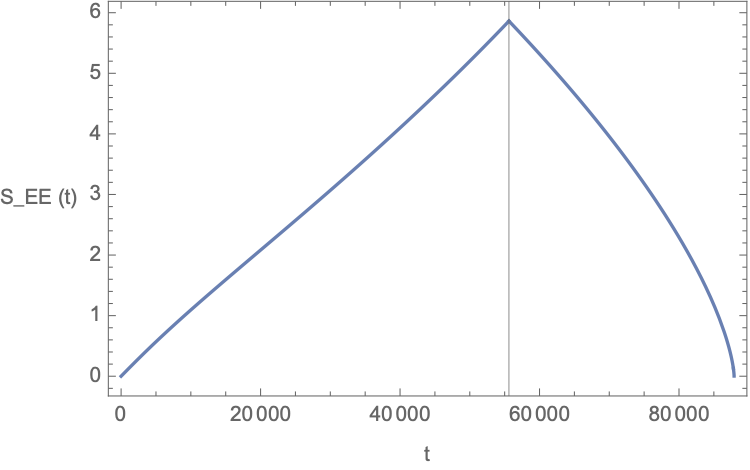}
        \caption{The semi-classical result of the Page curve for a Kerr black hole with $\beta = 1$ and the initial conditions $M(0)=1$,  $J(0)=1/2$}
        \label{fig:SemiClassicalPageCurveKerr}
      \end{center}
    \end{figure}

The entropies $S_{BH}$ and $S_{rad}$ can also be computed using \eqref{eq:S_BH} and the parameter $\beta$,  and subsequently,  the Page curve of a Kerr black hole can be obtained via \eqref{eq:ClassicalPageCurve}.  In the calculation,  the Hawking radiation rate $\Gamma_{s \omega l m p}$ is given by $\Gamma$ in \eqref{eq:Gamma} for fixed quantum numbers $(s, \omega,  l,  m,  p)$,  which in principle should include all the (nearly) massless particles in nature (photon,  neutrino,  graviton,  etc.).  For example,  the Page curve with only photon radiations was considered in \cite{Nian:2019buz},  and the resulting Page curve is shown in Fig.~\ref{fig:SemiClassicalPageCurveKerr}.

Related to the discussions in the previous section,  the time evolution of the entanglement entropy is contributed only from hard radiations,  i.e.,  the Hawking radiation particles carry energy and momentum away from a black hole,  causing it to evolve into a new state with smaller values of mass and momentum.


\subsection{Microscopic Computation of Page Curve}

In the previous section,  we have seen that the Hawking radiations cause the increasing entanglement entropy between the black hole and the Hawking quanta.  However,  as we briefly illustrated in Sec.~\ref{sec:KerrWormhole} and \ref{sec:InnerHorizon},  there can be a time-delay effect taking place in the interior of the Kerr black hole,  and the ingoing Hawking quanta can be absorbed by the inner horizon of the Kerr black hole and re-emitted from the white hole side.  Consequently,  the observer at the future infinity should observe not only the increase of the entanglement entropy between the black hole and the Hawking quanta,  but after a time delay,  the observer should also experience a decrease of the entanglement entropy caused by the Hawking quanta emitted from the white hole side.

In this section,  we will consider this effect and compute the time evolution of the entanglement entropy between Hawking quanta and the black hole-white hole system observed at the future infinity,  i.e.,  the Page curve.  We expect to reproduce the semi-classical result of the previous subsection in a microscopic way.  However,  these two curves do not need to look the same because the semi-classical result merely provides an upper bound of the instantaneous entanglement entropy.

As our previous qualitative discussions suggest,  to compute the Page curve in this picture,  the crucial step is to determine the time delay between the outgoing Hawking quanta and the ingoing Hawking quanta when they reach the future infinity.  In other words,  we need to measure the tunneling time of a Hawking particle inside the Kerr wormhole.  Let us analyze this process carefully in this subsection.

The first problem is that the time $t$ in the Boyer-Lindquist coordinates cannot cover more than one patch in the Kerr black hole spacetime because the proper time $t$ at infinity diverges at horizons $r_\pm$.  This issue has been addressed by Boyer and Lindquist themselves in \cite{Boyer:1966qh}.  The resolution is to define the Eddington-Finkelstein coordinates $u$ and $v$ for a Kerr black hole,  which can cover two patches of the Kerr black hole spacetime:
\begin{align}
  u & \equiv t - r - \frac{r_+^2 + a^2}{r_+ - r_-}\, \textrm{log} |r - r_+| + \frac{r_-^2 + a^2}{r_+ - r_-}\, \textrm{log} |r - r_-|\, ,\\
  v & \equiv t + r + \frac{r_+^2 + a^2}{r_+ - r_-}\, \textrm{log} |r - r_+| - \frac{r_-^2 + a^2}{r_+ - r_-}\, \textrm{log} |r - r_-|\, ,
\end{align}
where $u$ and $v$ are the retarded and the advanced times,  respectively.  For the purpose of this paper,  we adapt these definitions and define the modified retarded and the modified advanced times as follows:
\begin{align}
  \tau_u & \equiv u + r = t - \frac{r_+^2 + a^2}{r_+ - r_-}\, \textrm{log} |r - r_+| + \frac{r_-^2 + a^2}{r_+ - r_-}\, \textrm{log} |r - r_-|\, ,\\
  \tau_v & \equiv v - r = t + \frac{r_+^2 + a^2}{r_+ - r_-}\, \textrm{log} |r - r_+| - \frac{r_-^2 + a^2}{r_+ - r_-}\, \textrm{log} |r - r_-|\, ,
\end{align}
which have been introduced in \cite{Roken:2015fja}.  The advantages of using $\tau_{u,  v}$ are two-folded.  First,  the constant-$\tau_u$ and constant-$\tau_v$ curves are spacelike.  Second,  at a fixed $r$ such as the future infinity where $r \to \infty$,  the time differences measured by $t$,  $\tau_u$,  and $\tau_v$ coincide,  i.e.,
\be
  (\Delta t)_{r \to \infty} = (\Delta \tau_u)_{r \to \infty} = (\Delta \tau_v)_{r \to \infty}\, .
\ee
Therefore,  the different times $t$,  $\tau_u$,  and $\tau_v$ can be used interchangeably for an observer at the future infinity.

Because two neighboring patches of the Kerr spacetime can be covered by one of the new times $\tau_{u,  v}$,  we can measure the time delay in two neighboring patches. By gluing the patches,  we can obtain the tunneling time of a Hawking particle inside the two outer horizons $r_+$,  from the black hole side to the white hole side.  Let us analyze this process step by step.

{\flushleft $\bullet$ Step $\textrm{i}$:}

\begin{figure}[!htb]
      \begin{center}
        \includegraphics[width=0.41\textwidth]{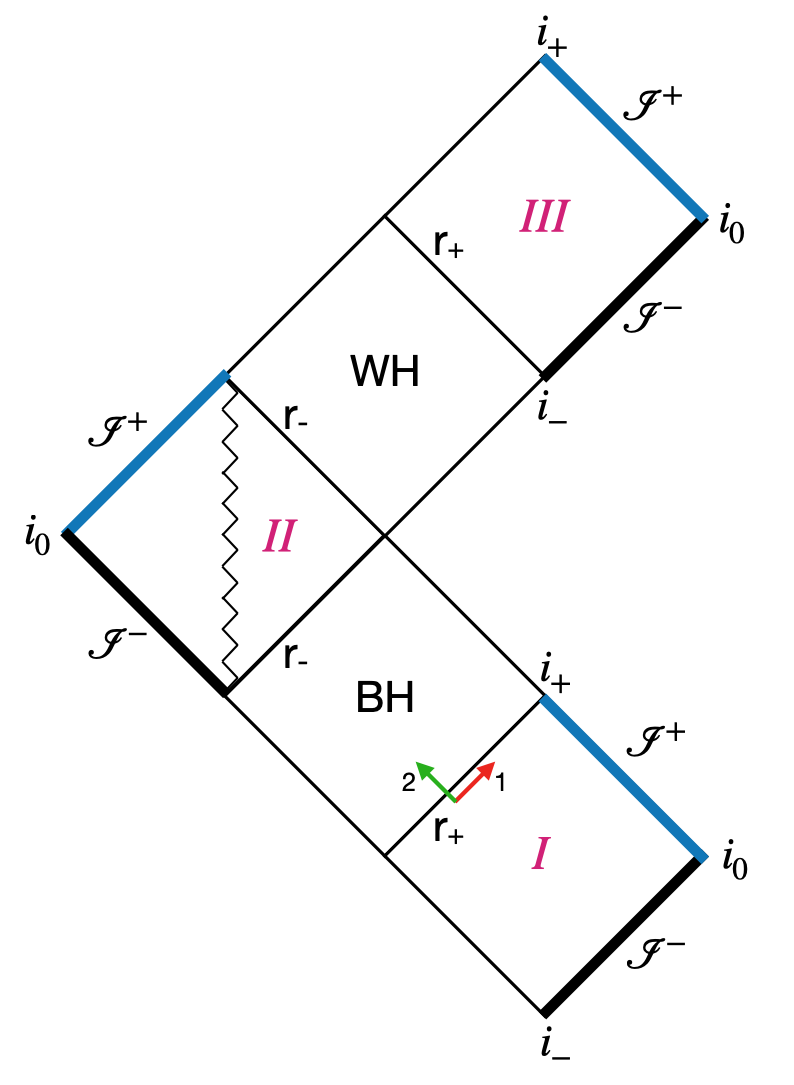}
        \caption{A pair of maximally entangled particles created at the outer horizon $r_+$}
        \label{fig:ParticlesAtOuterHorizonBH}
      \end{center}
    \end{figure}

After a pair of maximally entangled Hawking quanta $1$ and $2$ were created at the outer horizon $r_+$ of a Kerr black hole,  the particle $1$ will follow an outgoing lightlike geodesic to the future infinity (see Fig.~\ref{fig:ParticlesAtOuterHorizonBH}).

To compute the travel time of the particle $1$,  we can use the original Boyer-Lindquist coordinates,  and we suppress the angular coordinates by setting $\theta = \phi = 0$ for simplicity.  Consequently,  the travel time of the particle $1$ from the outer horizon $r_+$ to infinity is
\be
  (\Delta t)^{\text{Step }\textrm{i}}_{r \to \infty} = \int_{r_+}^\infty dr\, \left(\frac{dt}{dr} \right) = \int_{r_+}^\infty dr\, \frac{r^2 + a^2}{\Delta} = \int_{r_+}^\infty dr\, \left(1 + \frac{2 M r}{\Delta} \right)\, .
\ee
This integral expression is divergent.  In practice,  we can introduce a small scale $\delta$ and a finite large scale $r_\infty$ to regularize the divergence.  The regularized finite travel time of the particle $1$ is
\be
  (\Delta t)^{\text{Step }\textrm{i}}_{r \to \infty} = \int_{r_+ + \delta}^{r_\infty} dr\, \left(1 + \frac{2 M r}{\Delta} \right)\, .
\ee

\newpage

{\flushleft $\bullet$ Step $\textrm{i}'$:}

In contrast to its entanglement partner $1$,  the particle $2$ will follow a completely different path before eventually meeting particle $1$ again at the future infinity.  We divide the path of the particle $2$ into four parts,  called Step $\textrm{i}'$ through Step $\textrm{iv}'$.  Let us discuss them one by one.

After the particle $2$ enters the outer horizon,  the first stage (Step $\textrm{i}'$) is to travel from the outer horizon $r_+$ to the inner horizon $r_-$ of the black hole.  Measured by $t$,  it takes the following amount of time \cite{Roken:2015fja}:
\be
  (\Delta t)^{\text{Step }\textrm{i}'}_{r \to \infty} = \int_{r_+}^{r_-} dr\, \left(\frac{d\tau_v}{dr} \right)_{\text{in}} = \int_{r_+}^{r_-} dr\, (-1) = r_+ - r_-\, .
\ee
Hence,  in the natural units,  the travel time of the particle $2$ from the outer horizon $r_+$ to the inner horizon $r_-$ is
\be
  (\Delta t)^{\text{Step }\textrm{i}'}_{r \to \infty} = r_+ - r_-\, .
\ee

   \begin{figure}[!htb]
      \begin{center}
        \includegraphics[width=0.41\textwidth]{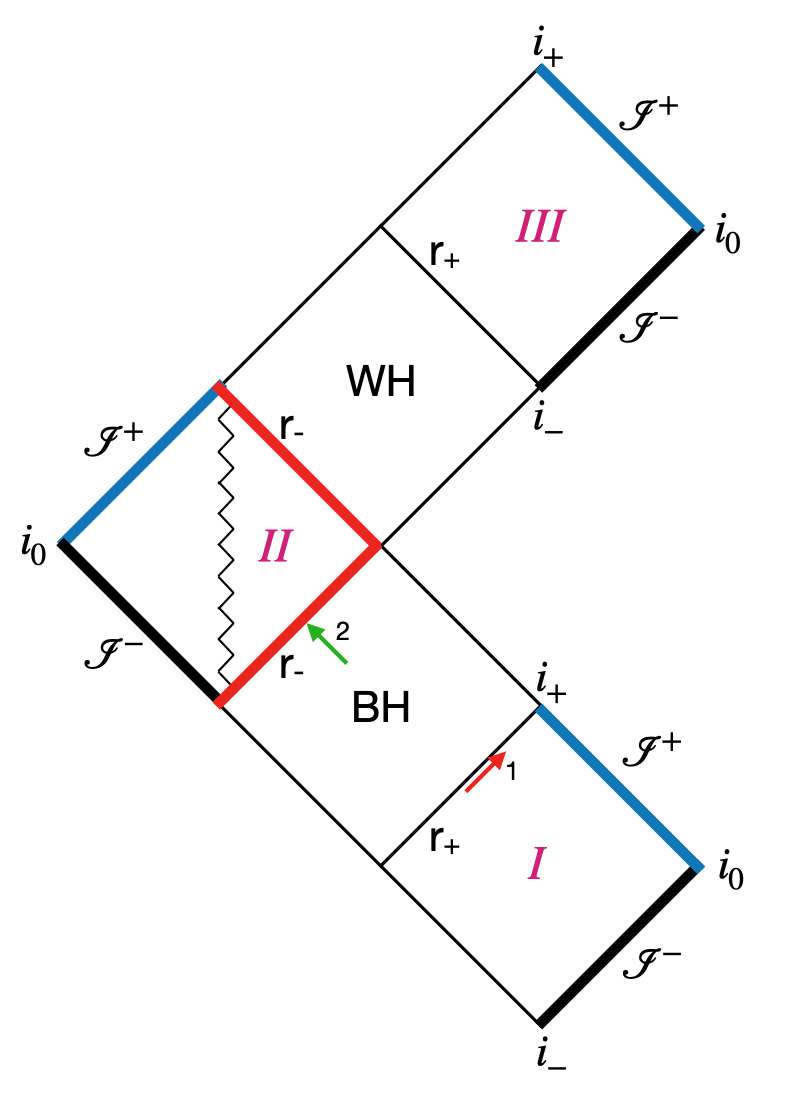}
        \caption{The particle $2$ arrives at the inner horizon $r_-$ on the black hole side}
        \label{fig:Step ii prime 1}
      \end{center}
    \end{figure}

\newpage

{\flushleft $\bullet$ Step $\textrm{ii}'$:}

After the particle $2$ arrives at the inner horizon (see Fig.~\ref{fig:Step ii prime 1}),  what will happen next? In classical gravity theory,  the inner horizon of a Kerr black hole is a Cauchy horizon.  Hence,  no precise theoretical prediction can be made beyond the inner horizon.  In addition,  the inner horizon may obey the strong cosmic censorship and consequently impose some strong constraints on the model.  Hence,  a concrete description of the physics beyond the inner horizon relies on the precise knowledge of quantum gravity theory,  which is beyond the scope of the current work.

\begin{figure}[!htb]
      \begin{center}
        \includegraphics[width=0.41\textwidth]{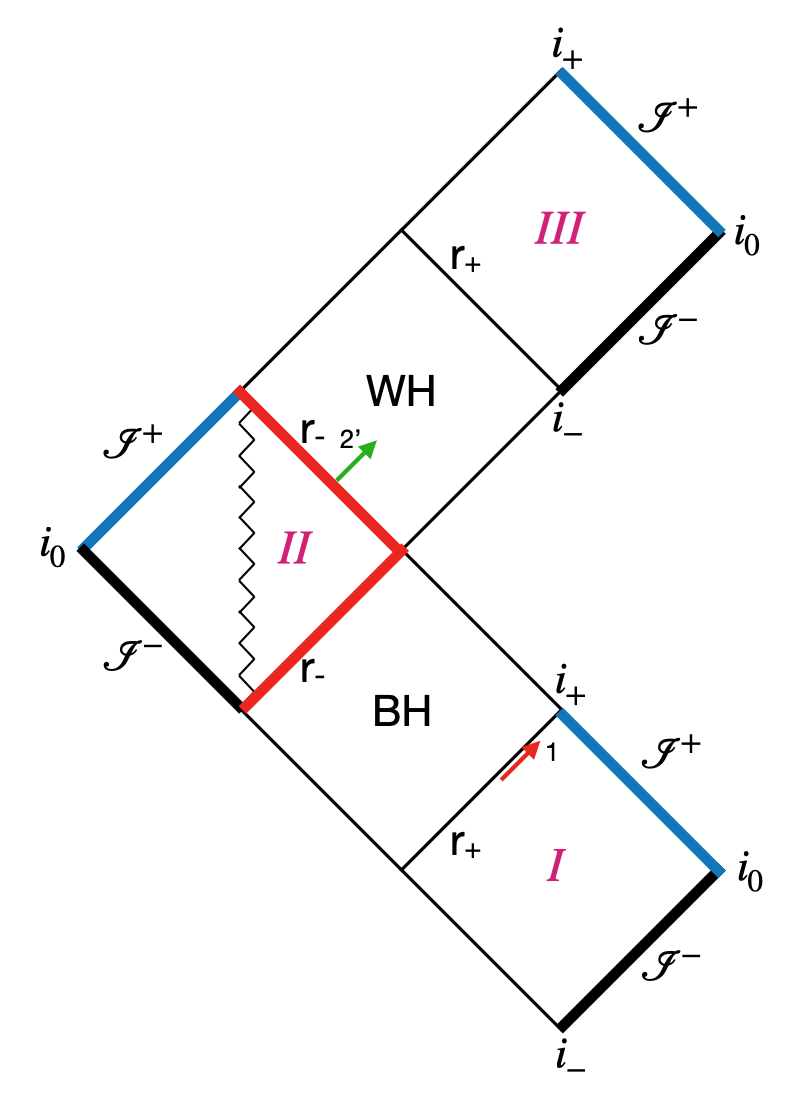}
        \caption{The particle $2'$ is emitted at the inner horizon $r_-$ on the white hole side}
        \label{fig:Step ii prime 2}
      \end{center}
    \end{figure}

In this paper,  we adopt a more pragmatic perspective,  generalizing the historical viewpoint of treating a black hole as an atom \cite{Bekenstein:1997bt,  EmparanSachs}.  More precisely,  we assume that the particle $2$ is absorbed by the inner horizon and its interior. Then, after a time delay, a new particle $2'$ will be emitted by the inner horizon of the white hole side (see Fig.~\ref{fig:Step ii prime 2}).  This new particle $2'$ maintains the maximal entanglement with the particle $1$.  In other words,  we treat the inner horizon with its interior as a ``giant'' atom.  The goal is to compute the time delay between the absorption and the emission,  just like the lifetime of a particular state in atomic physics.

\begin{figure}[!htb]
      \begin{center}
        \includegraphics[width=0.97\textwidth]{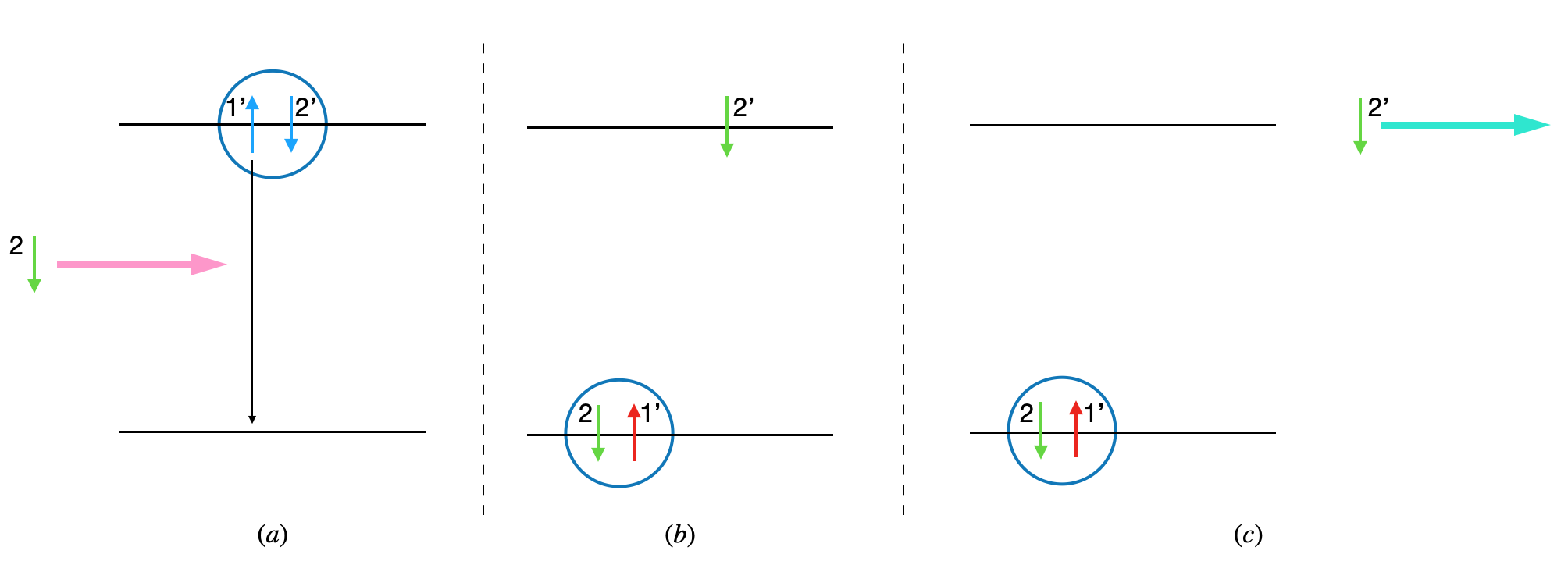}
        \caption{A schematic picture of the absorption and emission process}
        \label{fig:absorption and emission}
      \end{center}
    \end{figure}

With this picture in mind,  let us elaborate on the model in more detail.  First of all,  our new model is consistent with the models by Parikh and Wilczek \cite{Parikh:1999mf} and by Wen \cite{Wen:2019bjp},  where the Hawking radiation was treated as a tunneling process or a stimulated emission,  respectively.  The absorption and emission process can be schematically understood as shown in Fig.~\ref{fig:absorption and emission}.  In this process,  all the particles ($2$,  $1'$,  and $2'$) should be understood as quantum states,  or qubits,   while the spin-up and spin-down in Fig.~\ref{fig:absorption and emission} are just a schematic demonstration.  We will elaborate on this point later in Sec.~\ref{sec:QuInfo} and show that this absorption and emission process can be formulated as a quantum teleportation.

The whole absorption and emission process consists of three steps:
\begin{enumerate}
\item[(a)] Initially,  the black hole has $N(\omega, \, t_0)$ maximally entangled pairs of particles at the excited state with energy $\omega$.  At the initial time $t_0$,  the particle $2$ with energy $- \omega$ was absorbed by the inner horizon and its interior,  which triggers one particle $1'$ with energy $\omega$ of the black hole states to leap to the ground state.  Similar to \cite{Parikh:1999mf,  Wen:2019bjp},  the Hawking radiation particle $1$ can be understood as a stimulated emission of the particle $1'$ from the excited state to the ground state.

\item[(b)] The particles $2$ and $1'$ form a new maximally entangled pair on the ground state so that the total spin of the ground state remains unchanged.

\item[(c)] Next,  a spontaneous emission immediately follows the stimulated emission.  The remaining particle $2'$ with energy $\omega$ is spontaneously emitted by the inner horizon on the white hole side.  It carries energy away from the black hole-white hole system and maintains the maximal entanglement with the Hawking particle $1$ emitted from the black hole side. 
\end{enumerate}

Effectively,  this absorption and emission process happens at any fixed value of $\omega$ until the black hole-white hole system completely evaporates.  The emission rate from the white hole side can be computed from the sum of the production rates of the particle $2'$ over all the $\omega$'s.  For a fixed $\omega$,  due to the one-to-one correspondence between the initial and the final particles,  the total particle number emitted from the white hole side equals the total Hawking particle number emitted from the black hole side.

Now,  let us estimate the time delay between the absorption of particle $2$ and the emission of particle $2'$,  which is, on average, the lifetime of the particle $2'$ at the excited state.  As in atomic physics,  we can compute this lifetime associated with a spontaneous emission as follows:
\be
  (\Delta t)_{r \to \infty} (\omega,\, t) = \frac{N_{rem} (\omega,\, t)}{\frac{dN}{dt} (\omega,\, t)}\, ,
\ee
where $N_{rem} (\omega,\, t)$ denotes the remaining excited state number of energy $\omega$ at the time $t$,  and $\frac{dN}{dt} (\omega,\, t)$ is given by Eq.~\eqref{eq:dNdtPhoton} without the integral over $\omega$:
\be
  \frac{dN}{dt} (\omega,\, t) = \frac{1}{2 \pi} \sum_{mp} \langle N_{1w1mp} \rangle = \frac{1}{2 \pi} \sum_{mp}\, \Gamma_{1 \omega 1 m p}\, \frac{1}{\textrm{exp} \left[\frac{1}{T_H} (\omega - m \Omega) \right] - 1}\, .
\ee
By integrating $\frac{dN}{dt} (\omega,\, t)$ over time $t$,  we obtain the total excited state number $N_{tot} (\omega)$ at a given $\omega$,  and the remaining excited state number $N_{rem} (\omega,\, t)$ at time $t$ is
\be
  N_{rem} (\omega,\, t) = N_{tot} (\omega) - \int_0^t d\tilde{t}\,\, \frac{dN}{dt} (\omega,\, \tilde{t})\, .
\ee

\begin{figure}[!htb]
      \begin{center}
        \includegraphics[width=0.88\textwidth]{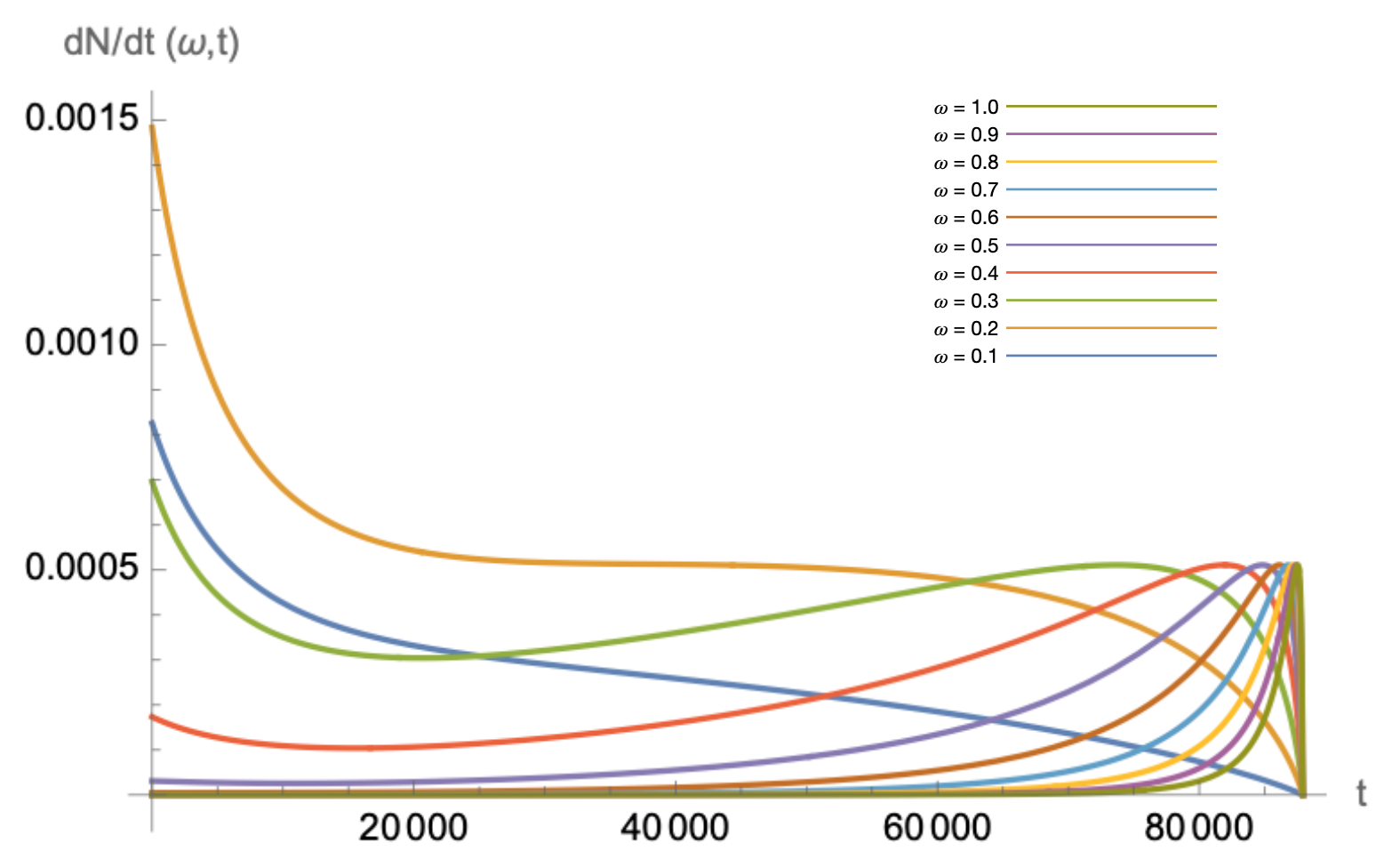}
        \caption{$\frac{dN}{dt} (\omega,\, t)$ for different values of $\omega$}
        \label{fig:dNdt omega}
      \end{center}
    \end{figure}

\begin{figure}[!htb]
      \begin{center}
        \includegraphics[width=0.67\textwidth]{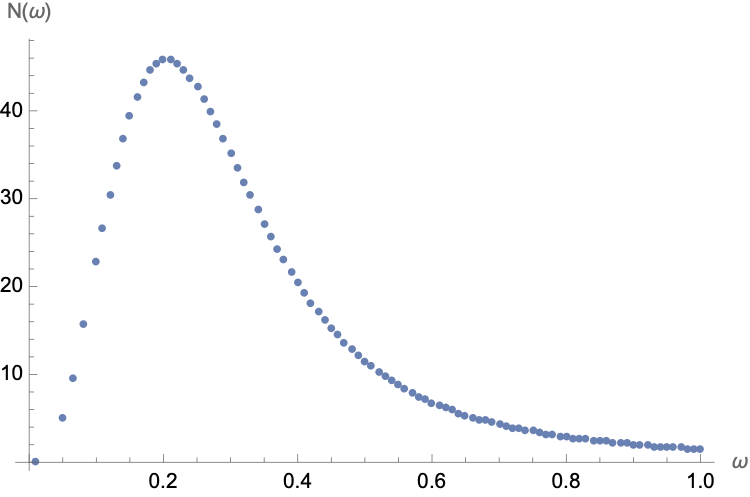}
        \caption{$N (\omega)$ for different values of $\omega$}
        \label{fig:N omega}
      \end{center}
    \end{figure}

Noticing that $\frac{dN}{dt}$ is composed of different $\omega$'s contributions,  we can analyze the components of $\frac{dN}{dt}$ and already see some interesting features at this stage.  For instance,  we can plot for the given initial values ($M(0)=1$,  $J(0)=1/2$) the time-dependences of $\frac{dN}{dt} (\omega,\, t)$ (see Fig.~\ref{fig:dNdt omega}) and the total excited state number $N_{tot} (\omega)$ (see Fig.~\ref{fig:N omega}) with different values of $\omega$.

\newpage
From these figures,  we see that the emission rates at different $\omega$'s have quite different behaviors,  and the most probable energy of emitted particles is around $\omega = 0.2$ for the initial values $M(0)=1$,  $J(0)=1/2$.

{\flushleft $\bullet$ Step $\textrm{iii}'$:}

\begin{figure}[!htb]
      \begin{center}
        \includegraphics[width=0.41\textwidth]{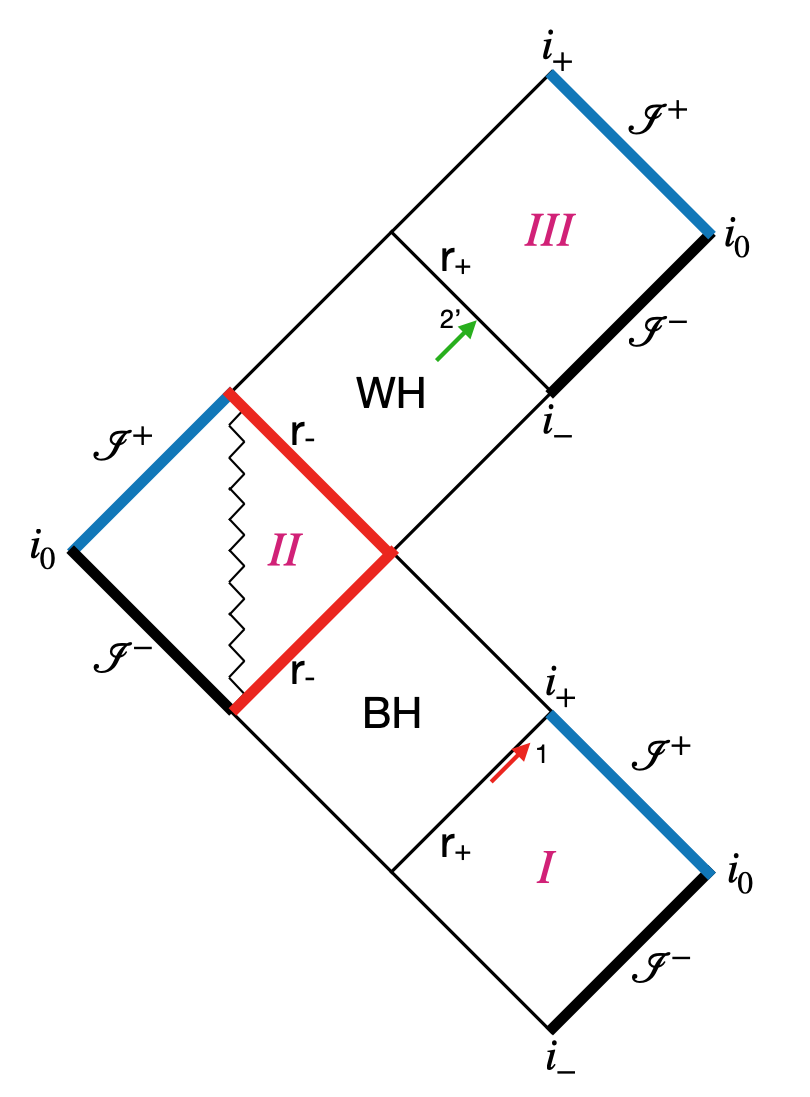}
        \caption{The particle $2'$ arrives at the outer horizon $r_+$ on the white hole side}
        \label{fig:Step iii prime}
      \end{center}
    \end{figure}

After the new particle $2'$ emitted from the inner horizon $r_-$ on the white hole side,  it travels from the inner horizon $r_-$ to the outer horizon $r_+$ (see Fig.~\ref{fig:Step iii prime}),  taking the same amount of time as Step $\textrm{i}'$:
\be
  (\Delta t)^{\text{Step }\textrm{iii}'}_{r \to \infty} = r_+ - r_-\, .
\ee
This can be computed in a way similar to Step $\textrm{i}'$,  but using the modified retarded time $\tau_u$.

{\flushleft $\bullet$ Step $\textrm{iv}'$:}

\begin{figure}[!htb]
      \begin{center}
        \includegraphics[width=0.41\textwidth]{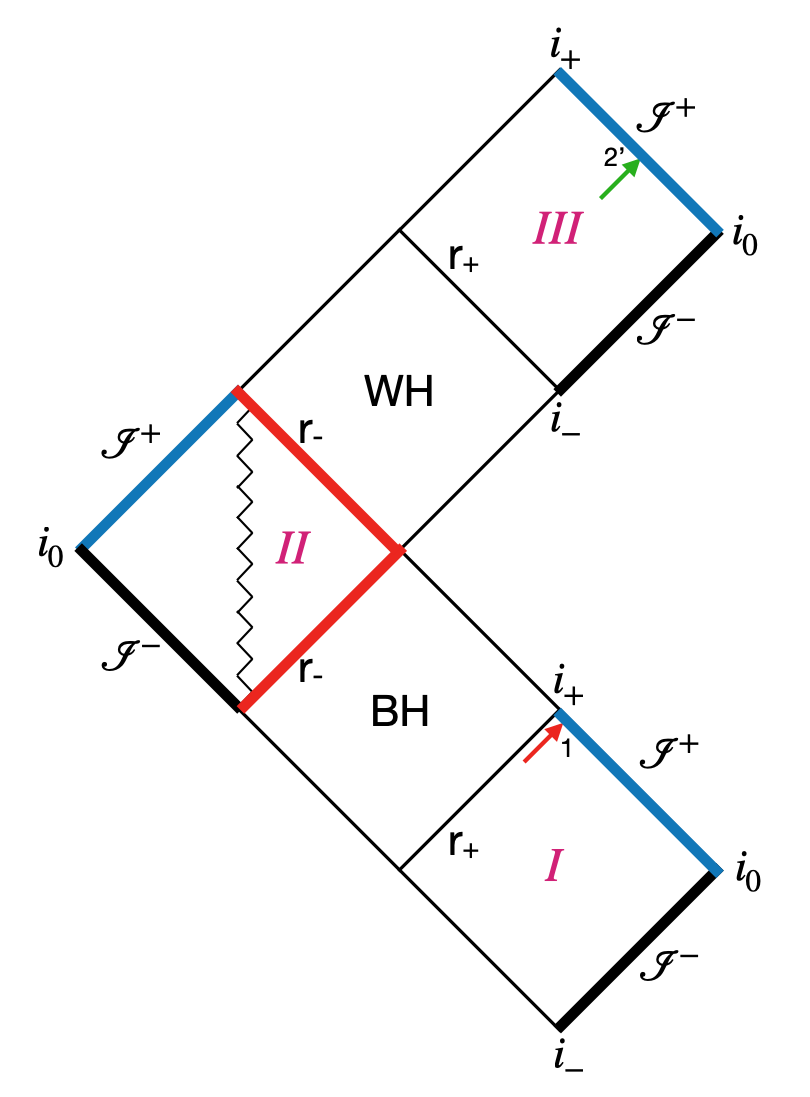}
        \caption{The particles $1$ and $2'$ both arrive at the future infinity $\mathscr{I}^+$ with a time difference}
        \label{fig:Step iv prime}
      \end{center}
    \end{figure}

In this step,  the particle $2'$ is an outgoing particle traveling from the outer horizon $r_+$ of the white hole to the future infinity $\mathscr{I}^+$ (see Fig.~\ref{fig:Step iv prime}).  Therefore,  the particles $1$ and $2'$ meet at the future infinity but with a time delay,  and the entanglement is purified in the final state.

Similar to Step $\textrm{i}$ of the particle $1$,  we can use the original Boyer-Lindquist coordinates in this patch,  and after regularization, the finite travel time of the particle $2'$ is
\be
  (\Delta t)^{\text{Step }\textrm{iv}'} = \int_{r_+ + \delta}^{r_\infty} dr\, \left(1 + \frac{2 M r}{\Delta} \right)\, ,
\ee
which is equal to the travel time of the particle $1$ in Step $\textrm{i}$.

At this moment,  let us make a summary of the current results:
\begin{itemize}
\item[-] The particle $1$ in Step $\textrm{i}$ and the particle $2'$ in Step $\textrm{iv}'$ have the equal (divergent) travel time.  In practice,  introducing cutoffs near the outer horizon and at infinity can regularize the divergence and lead to a finite travel time.

\item[-] Hence,  the total time difference between the maximally entangled Hawking pair receives contributions only from Steps $\textrm{i}'$,  $\textrm{ii}'$,  $\textrm{iii}'$.

\item[-] The travel time of the particle $2$ in Step $\textrm{i}'$ and the one of the particle $2'$ in Step $\textrm{iii}'$ are both $(r_+ - r_-)$.

\item[-] The delay in Step $\textrm{ii}'$ is the particle lifetime at an excited state of a black hole if the inner horizon with its interior is treated as an atom.  Inspired by atomic physics,  the lifetime can be computed as $N_{rem} \cdot (dN/dt)^{-1}$ for a given frequency $\omega$,  where $\frac{dN}{dt} (\omega,\, t)$ is just the black hole's Hawking radiation rate at energy $\omega$,  inspired by the approach of viewing Hawking radiations as stimulated emissions \cite{Parikh:1999mf,  Wen:2019bjp}.

\item[-] When two maximally entangled Hawking quanta $1$ and $2'$ both reach the future infinity $\mathscr{I}^+$,  their arrival time differs by the total amount
\begin{align}
  (\Delta t)_{r \to \infty} & = (\Delta t)^{\text{Step }\textrm{i}'}_{r \to \infty} + (\Delta t)^{\text{Step }\textrm{ii}'}_{r \to \infty} + (\Delta t)^{\text{Step }\textrm{iii}'}_{r \to \infty} \nonumber\\
  {} & = (\Delta \tau_v)^{\text{Step }\textrm{i}'}_{r \to \infty} + (\Delta t)^{\text{Step }\textrm{ii}'}_{r \to \infty} + (\Delta \tau_u)^{\text{Step }\textrm{iii}'}_{r \to \infty} \nonumber\\
  {} & = (r_+ - r_-) + N_{rem}\cdot \left(\frac{dN}{dt}\right)^{-1} + (r_+ - r_-) \nonumber\\
  {} & = 2\, (r_+ - r_-) + N_{rem}\cdot \left(\frac{dN}{dt}\right)^{-1}\, .\label{eq:TimeDelay}
\end{align}
From the previous results,  we know that $r_\pm$ are functions in $t$,  while $dN/dt$ depends on $(\omega,\, t)$.  For an outgoing Hawking particle $1$ emitted from the black hole side,  when it reaches the observer at the infinity,  we anticipate the arrival of its entangled partner $2'$ from the white hole side after a time delay $(\Delta t)_{r \to \infty}$.  This means that the created entanglement will be purified after a certain time delay.  A similar setup of computing the tunneling time through a Kerr-like geometry has been employed in \cite{Han:2023wxg} in a different context.

$\quad$ An interesting fact is that for a generic non-extremal Kerr black hole,  the delay time given by Eq.~\eqref{eq:TimeDelay} is nonzero when the black hole evaporation starts,  but eventually it vanishes when the black hole completely evaporates.  This can be seen from the final values of $r_+$,  $r_-$,  and $N_{rem}$,  which all vanish at the end of the evaporation.

\end{itemize}

Finally,  based on all the considerations above,  we can compute the radiation rate from the white hole side,  which is proportional to the one from the black hole side with a time delay given by \eqref{eq:TimeDelay},  i.e.,
\be\label{eq:dNdtWH}
  \left(\frac{dN}{dt} \right)^{WH} (\omega,\, t) = \mathcal{N}\cdot \left(\frac{dN}{dt} \right)^{BH} \left(\omega,\, t + (\Delta t)_{r \to \infty} \right)\, ,
\ee
where $\mathcal{N}$ is a normalization factor to guarantee that the total emitted particle numbers from the black hole and the white hole sides match.  The expression \eqref{eq:dNdtWH} is the key step of this paper.

\begin{figure}[!htb]
      \begin{center}
        \includegraphics[width=0.67\textwidth]{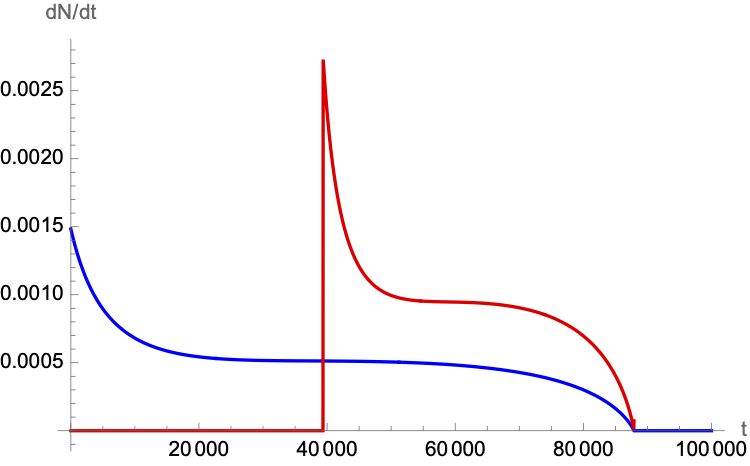}
        \caption{$\left(\frac{dN}{dt} \right)^{BH} (\omega,\, t)$ (blue curve) and $\left(\frac{dN}{dt} \right)^{WH} (\omega,\, t)$ (red curve) for $\omega=0.2$ and the initial values $M(0)=1$,  $J(0)=1/2$}
        \label{fig:dNdtBHandWH}
      \end{center}
    \end{figure}

With the above results,  we can numerically compute the radiation rates with very high precision.  An prototype of the curves $\left(\frac{dN}{dt} \right)^{BH} (\omega,\, t)$ and $\left(\frac{dN}{dt} \right)^{WH} (\omega,\, t)$ is shown in Fig.~\ref{fig:dNdtBHandWH}.

\begin{figure}[!htb]
      \begin{center}
        \includegraphics[width=0.67\textwidth]{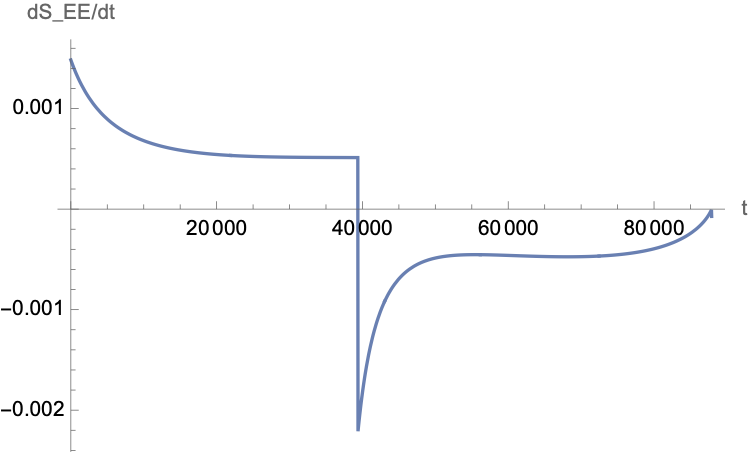}
        \caption{$\frac{dS_{EE}}{dt} (\omega,\, t)$ for $\omega=0.2$ and the initial values $M(0)=1$,  $J(0)=1/2$}
        \label{fig:dSdt}
      \end{center}
    \end{figure}

\newpage
Again,  the radiations from the black hole side cause the increase of the entanglement between Hawking quanta and the black hole-white hole system,  while the radiations from the white hole side cause the decrease of the entanglement.  Taking the difference of $\left(\frac{dN}{dt} \right)^{BH} (\omega,\, t)$ and $\left(\frac{dN}{dt} \right)^{WH} (\omega,\, t)$ shown in Fig.~\ref{fig:dNdtBHandWH},  we obtain the change of the entanglement in time,  i.e.,  $\frac{dS_{EE}}{dt} (\omega,\, t) = \left(\frac{dN}{dt} \right)^{BH} (\omega,\, t) - \left(\frac{dN}{dt} \right)^{WH} (\omega,\, t)$ shown in Fig.~\ref{fig:dSdt},  where we drop a factor $\textrm{log} (2)$ in $\frac{dS_{EE}}{dt}$.

\begin{figure}[!htb]
      \begin{center}
        \includegraphics[width=0.67\textwidth]{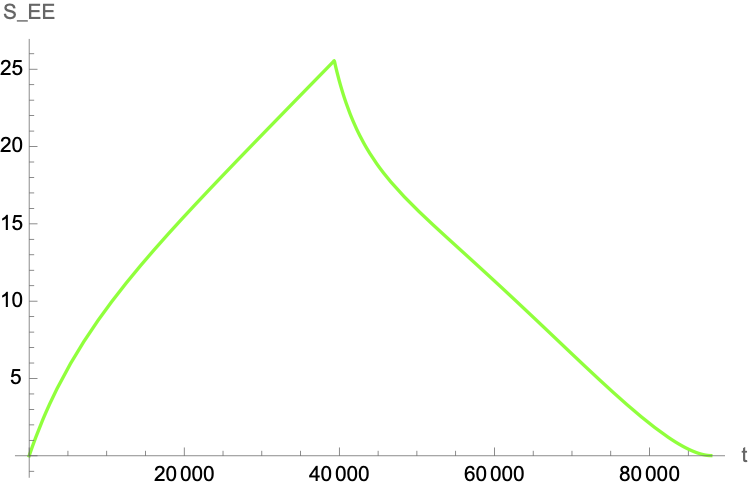}
        \caption{$S_{EE} (\omega,\, t)$ for $\omega=0.2$ and the initial values $M(0)=1$,  $J(0)=1/2$}
        \label{fig:SEE at w=0.2}
      \end{center}
    \end{figure}

\newpage
Integrating the numerical results of $\frac{dS_{EE}}{dt} (\omega,\, t)$ shown in Fig.~\ref{fig:dSdt},  we obtain the time evolution of the entanglement entropy between the Kerr black hole and its Hawking radiation quanta,  i.e.,   the Page curve at a given frequency $\omega$,  shown in Fig.~\ref{fig:SEE at w=0.2}.

We see that Fig.~\ref{fig:SEE at w=0.2} meets all the expectations from the semi-classical result,  shown in Fig.~\ref{fig:SemiClassicalPageCurveKerr},  and it is significantly different compared to the classical result following Hawking's approach,  shown in Fig.~\ref{fig:SEEfromBH}.  Instead of monotonical increasing and being saturated at a maximal value,  $S_{EE} (\omega,\, t)$ from the microscopic approach reaches the maximum at the Page time and then decreases to zero,  which indicates a pure state for the final state of black hole evaporation.

\begin{figure}[!htb]
      \begin{center}
        \includegraphics[width=0.67\textwidth]{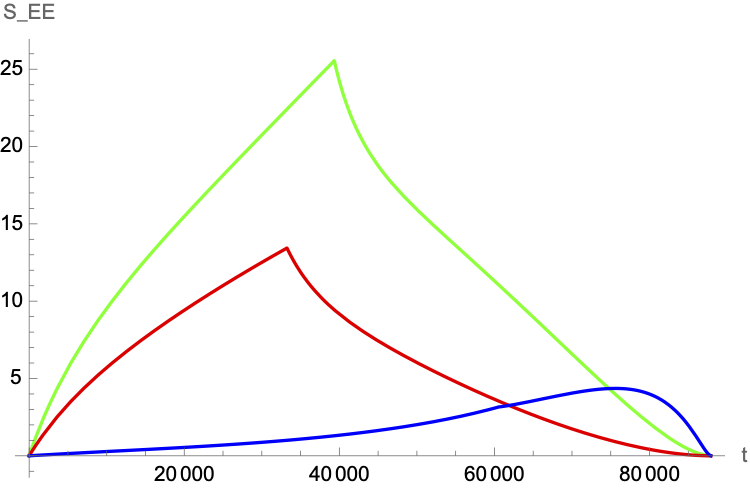}
        \caption{$S_{EE} (\omega,\, t)$ for $\omega=0.1$ (red curve),  $\omega=0.2$ (green curve),  and $\omega=0.5$ (blue curve) with the initial values $M(0)=1$,  $J(0)=1/2$}
        \label{fig:SEE at 3 omegas}
      \end{center}
    \end{figure}

However,  this is not the end of the story.  Remember that we have considered only one value of $\omega$,  but Hawking radiations contain all positive frequencies with different values of $\omega$.  The final result of total entanglement entropy $S_{EE} (t)$ receives contributions from all $\omega$'s,  which requires a careful ensemble average.  As the first step,  we compute and compare the results for a few values of $\omega$.  As we see in Fig.~\ref{fig:N omega},  $\omega=0.2$ has the maximal value of particle radiation number for the initial values $M(0)=1$,  $J(0)=1/2$.  To make a comparison,  we compute $S_{EE} (\omega,\, t)$ for $\omega = 0.1$,  $0.2$,  $0.5$,  and plot all three curves in Fig.~\ref{fig:SEE at 3 omegas}.

All the $S_{EE} (\omega,\, t)$ curves have similar features.  They increase from zero until a peak, then decrease to zero at the end.  Hence,  a pure state always evolves into a pure state,  which holds at all frequencies $\omega$.  The main difference is the position of the turning point,  i.e.,  the Page time.  For $\omega=0.2$,  the Page time is roughly half of the black hole evaporation time,  but for other values of $\omega$, the Page time can be earlier or later.  Again,  the precise determination of the Page time requires a careful ensemble average over all $\omega$'s,  which is left to future research.  Qualitatively,  $\omega=0.2$ provides the maximal radiation particle number. Hence,  the Page curve for $\omega=0.2$ can be viewed as a saddle-point approximation of the genuine Page curve with the actual Page time.  Comparing it with the semi-classical result shown in Fig.~\ref{fig:SemiClassicalPageCurveKerr},  we see that the microscopic result meets all the expectations from the semi-classical computation.

In this paper,  the Page time has a clear physical meaning.  This is the time when the inner horizon on the white hole side starts to emit particles,  which induces a ``kink'' in the Page curve.  In other words,  this is when the inner horizon becomes important.  Hence,  in some sense,  the inner horizon in our model plays a similar role as the island in the recent works \cite{Penington:2019npb,  Almheiri:2019psf,  Penington:2019kki,  Almheiri:2019qdq}.

\begin{figure}[!htb]
      \begin{center}
        \includegraphics[width=0.67\textwidth]{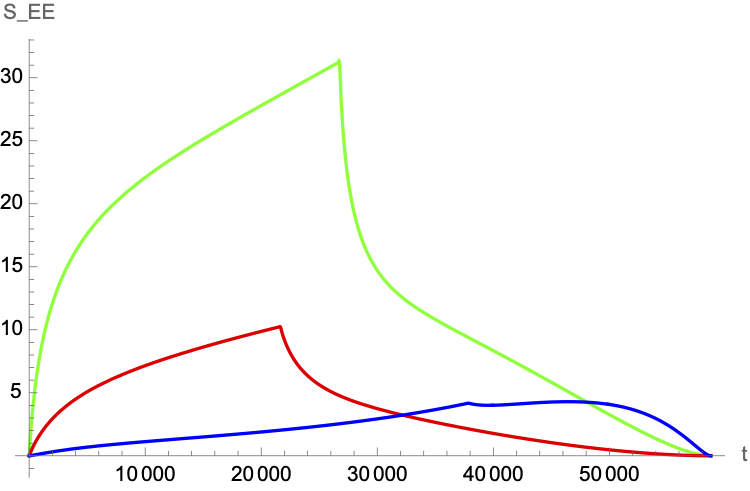}
        \caption{$S_{EE} (\omega,\, t)$ for $\omega=0.1$ (red curve),  $\omega=0.24$ (green curve),  and $\omega=0.5$ (blue curve) with the initial values $M(0)=1$,  $J(0)=0.95$}
        \label{fig:SEE at 3 omegas near extremal}
      \end{center}
    \end{figure}

We also change initial values and repeat the same steps to compute the corresponding Page curves.  For instance,  we consider $M(0) = 1$ and $J(0)=0.95$,  which is closer to the extremality.  The results at different values of $\omega$ are shown in Fig.~\ref{fig:SEE at 3 omegas near extremal}.  We see that the Page curves maintain the same qualitative features,  i.e.,  they start from zero,  end at zero,  and take peak values at some intermediate Page time.  Again,  the genuine Page curve can be obtained as an ensemble average of the curves at different $\omega$'s,  which should be dominated by the saddle-point approximation (the curve with $\omega = 0.24$ in this case).  Hence,  we conclude that our approach is robust,  ranging from near-extremal to generic non-extremal Kerr black holes.

Besides the recently proposed method based on replica wormholes and the island formula for the Euclidean Schwarzschild black holes \cite{Penington:2019npb,  Almheiri:2019psf,  Penington:2019kki,  Almheiri:2019qdq},  there are also other proposals in the literature (e.g.,  \cite{Zeng:2021kyb,  Zeng:2022brh,  Chu:2022ieq,  Chu:2023agv,  Feng:2023pfq}) trying to derive the Page curves.  However,  none of these previous attempts have included the rotation of black holes.  This paper shows that rotation is crucial in modifying the spacetime structure, providing a resolution with precise physical meanings to the information paradox.

So far,  we have seen that the white hole side also emits particles,  carrying away energy and angular momentum.  This raises an interesting question: What does the black hole mass $M$ really mean? In Page's model of black hole evaporation,  the black hole mass $M$ is eventually completely radiated.  Our results suggest that the black hole radiated mass might be just a portion of the total mass of the black hole-white hole system.  A detailed study on the energy balance is left for future research.  In this paper,  we take $M$ as the total energy carried away by Hawking radiation quanta from the black hole side.

\subsection{Interpretation in Quantum Information Theory}\label{sec:QuInfo}

Now,  let us return to the key step of the absorption and emission process at the inner horizon.  As mentioned before,  the absorption and emission process shown in Fig.~\ref{fig:absorption and emission} is just schematic.  All the spin-up and spin-down in that figure should be understood as quantum states,  or qubits,  instead of classical spin configurations.  More precisely,  the process shown in Fig.~\ref{fig:absorption and emission} should be formulated in a quantum information language.  This indeed can be done.  Surprisingly, this process inside a black hole precisely mimics the quantum teleportation process.

\begin{figure}[!htb]
      \begin{center}
        \includegraphics[width=0.95\textwidth]{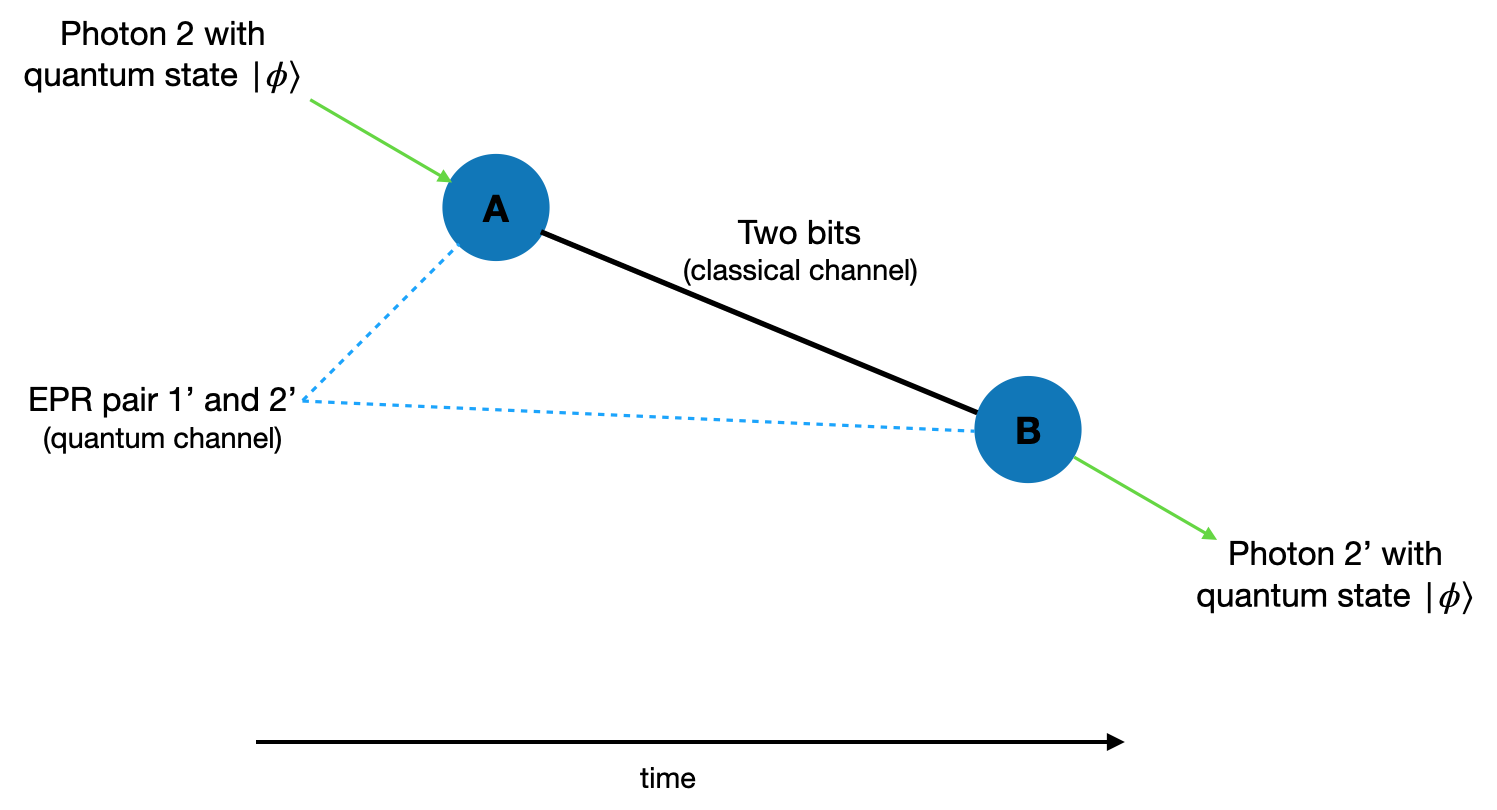}
        \caption{A standard quantum teleportation protocol is similar to the absorption and emission process at the inner horizon.}
        \label{fig:Qu Teleportation}
      \end{center}
    \end{figure}

In quantum information theory,  there is a well-known theorem,  the no-cloning theorem, which says that a generic quantum state (or a qubit) cannot be copied through unitary operations \cite{QuInfoBook}.  However,  a qubit can be teleportated,  i.e.,  it can be first destroyed at one point and then reappear at another point in spacetime through a unitary quantum-mechanical process.  A standard quantum teleportation protocol is shown in Fig.~\ref{fig:Qu Teleportation}.  To teleport a quantum state (e.g.,  a photon $2$),  we need another EPR pair,  i.e.,  a pair of maximally entanglement photons ($1'$ and $2'$).  After using the photons $2$ and $1'$ to perform a Bell measurement at Point $A$,  the measurement results can be stored in two classical bits and transferred to Point $B$,  which is not instantaneous and takes some time.  According to the two-bit measurement results,  the quantum state $2$ can be reconstructed by performing some unitary operations on the other photon $2'$ in the EPR pair \cite{QuInfoBook}.


In the quantum teleportation process described above,  we deliberately used the same names of the photons as in the absorption and emission process sketched in Fig.~\ref{fig:absorption and emission}.  Moreover,  the transfer time of the classical bits from Point $A$ to Point $B$ corresponds to the delay time of Step $\textrm{ii}'$ inside $r_-$,  i.e.,  $(\Delta t)^{\text{Step }\textrm{ii}'}_{r \to \infty}$.  Hence,  we see that what happens inside the inner horizon of a Kerr black hole can indeed be interpreted as a quantum teleportation.  In other words,  the inner horizon with its interior teleports the incoming Hawking particle $2$ from the black hole side to the white hole side,  and the new particle $2'$ maintains its maximal entanglement with the outgoing particle $1$ emitted from the black hole side.  Therefore,  the whole picture of our model has been put on the solid ground of quantum information theory.


\section{Discussion}\label{sec:Discussion}

In this paper,  we study the entanglement entropy between a Kerr black hole and its Hawking radiation quanta during evaporation.  We revisited the gravity approach for computing the Hawking radiation rate and formulated it as an equivalent CFT calculation.  From the results,  we see how the information paradox emerges.  To resolve it,  we computed the Page curve using a semi-classical and a microscopic approach.  To compute the Page curve microscopically,  we carefully keep track of the ingoing Hawking particle through the Kerr black hole and find that the Page curve can be understood as a time-delay effect.  The Page curves from both approaches match each other,  supporting the consistency of the new microscopic model.

As a highlight,  we find that the crucial step of the model,  i.e.,  the absorption and emission process at the inner horizon,  can be naturally interpreted as a quantum teleportation process,  which unvails the quantum nature of Kerr black holes.  Meanwhile,  the unitarity of the black hole evaporation process becomes manifest in the model,  which boils down to the unitarity of the absorption and emission process of the Hawking quanta at the inner horizon.

This paper provides a framework as a first step towards resolving the information paradox for Kerr black holes. There are many improvements to be made in the future.  For instance,  the genuine Page curve should be an ensemble average over all the frequencies,  but this paper has only considered a leading-order contribution.  A more delicate result summing over all the frequencies is desired.  More precisely,  each particle going through the wormhole can cause a slight change of the spacetime.  Based on the model in this paper,  we should try to establish a more refined model of the Kerr black hole evaporation.  In principle, a time-dependent dynamical simulation of the whole black hole evaporation process should be feasible.

There has been a long history of studying the black hole information paradox,  and many models have been proposed.  Although the new proposal in this paper differs significantly from all the previous works,  there are still natural connections.  Let us briefly comment on the literature.  This paper presents the apparent paradox for the evolution of entanglement entropy between Hawking's approach and Page's semi-classical approach.  In contrast,  some works are based on proposals other than the Page curve (e.g.,  \cite{Mathur:2005zp,  Ho:2022gpg}).  For those proposals trying to reproduce the Page curve,  almost all the existing works in the literature consider non-rotating black holes.  In this work,  we have demonstrated that the rotation of a black hole does not complicate the problem. Instead,  it provides some deformation, making the original problem more tractable.  Hence,  in our approach, the Schwarzschild black holes can be treated as Kerr black holes in the limit of zero angular momentum,  and the information paradox has already been resolved within the framework of Kerr black holes before taking the limit.

The recent progress based on the replica wormhole and the island formula is mainly for Schwarzschild black holes in Euclidean gravity \cite{Penington:2019npb,  Almheiri:2019psf,  Penington:2019kki,  Almheiri:2019qdq}.  This differs from the Kerr black holes in Lorentzian gravity considered in this paper.  However,  as we mentioned before,  the inner horizon in our model indeed plays a similar role as the island in \cite{Penington:2019npb,  Almheiri:2019psf,  Penington:2019kki,  Almheiri:2019qdq}.  Page time is the time when the inner horizon starts to re-emit particles,  comparable to the other picture when the island starts to dominate the entanglement entropy.  These interesting similarities suggest some deep connections between these different approaches.

The inner horizon of a Kerr black hole is expected to be a Cauchy horizon,  i.e., classical physics can be predicted from the initial data up to this point.  There has been a long history of studying the stability of the inner horizons (see,  e.g.,  \cite{McNamara}).  In particular,  the stability of the inner horizon as a Cauchy horizon is related to the strong cosmic censorship conjecture \cite{Penrose},  which states that a spacetime perturbation will destroy the horizon.  Although the strong cosmic censorship conjecture has been disproven to some extent \cite{Dafermos:2017dbw,  Dias:2018etb},  the stability of the inner horizon is still a concern in the resolution of the information paradox for Kerr black holes.  In this paper,  we abandon the classical description of the interior of the inner horizon,  which the strong cosmic censorship conjecture relies on.  Instead,  we argue that a quantum-mechanical description should be evoked to study the absorption and emission process inside the inner horizon.  Although this picture still needs further study,  it can provide qualitatively correct results and is supported by a quantum information interpretation.  Hence, instead of the classical properties of the inner horizon with its interior, the quantum ones should be more relevant to the information paradox.

Dray and 't Hooft found in \cite{Dray:1984ha} that a massless particle moving at the speed of light along the horizon of the Schwarzschild black hole can cause a gravitational shock wave,  which consequently leads to a shift of the horizon.  This effect is crucial in understanding the physical process of Hawking radiation.  In a more realistic model,  the Hawking radiations are not always spherical $s$-waves, particularly for the axially symmetric Kerr black holes. To emit one Hawking radiation particle,  the horizon will experience a recoil and deform a little.  However,  since a huge amount of such radiations happen randomly at each angle,  the recoil forces are balanced on average.  Hence,  roughly speaking,  the Kerr black hole will maintain its axial symmetry at the macroscopic scale,  but this picture deserves a more detailed study.

As we have seen in this paper,  the white hole plays a pivotal role in the new solution.  Is it an observable astrophysical phenomenon? Or is it just a theoretical construction? White holes have never been seriously studied in the literature,  partially because of the lack of actual observations.  However,  there are some potential candidates for white holes,  for instance,  the gamma-ray burst observed by NASA in 2006 \cite{Gehrels:2006tk}.

In this paper,  we found that during the black hole evaporation, both the temperature and the Hawking radiation rate have exponential growths towards the end evaporation (see Figs.~\ref{fig:T(t)} and \ref{fig:dNdt}).  We expect the same results to hold for the white hole side.  Hence,  due to the very high temperature and radiation rate at late time,  black and white holes can release all the remaining energy through gamma-ray bursts.  In one of his early papers on Hawking radiation \cite{BHexplosion},  Hawking suggested an explosion of a black hole as its final destiny.  Based on our preliminary numerical results,  we propose that a gamma-ray burst is possible as the fate of black and white holes.

Because a pair production from a vacuum can occur anywhere in spacetime due to vacuum fluctuations,  it may also happen in the vicinity of the inner horizon on the white hole side.  Effectively,  we will see Hawking radiations from the inner horizon of the white hole.  Does this effect cause a new information paradox? Yes,  but the new paradox can be resolved similarly using the approach introduced in this paper.  We hope to provide a thorough analysis of this novel phenomenon in the near future.

\section*{Acknowledgments}

The author would like to thank Bin Chen,  Xiaoyong Chu,  Wei Cui,  Xianhui Ge,  Sven Bjarke Gudnason,  Song He,  Xing Huang,  Qiang Jia,  Yi Ling,  Hong L\"u,  Leo A.  Pando Zayas,  Fernando Quevedo,  Wei Song,  Jia Tian,  Weijie Tian,  Juntao Wang,  Qiang Wen,  Jieqiang Wu,  Fengjun Xu,  Zhenbin Yang,  Jinwu Ye,  Yi Zhang,  and Yang Zhou for many helpful discussions.  Special thanks to Yu Tian and Hongbao Zhang for helping resolve some key issues in the paper.  This work is supported in part by the NSFC under grants No.~12375067 and No.~12147103.  The author would like to thank the International Centre for Theoretical Physics (ICTP) and the Korea Institute for Advanced Study (KIAS) for the warm hospitality during the final stage of this work.

\bibliographystyle{utphys}
\bibliography{KerrBHInfoParadox}

\providecommand{\href}[2]{#2}\begingroup\raggedright\begin{thebibliography}{10}

\bibitem{Hawking:1974sw}
S.~W. Hawking, ``{Particle Creation by Black Holes},''
\href{http://dx.doi.org/10.1007/BF02345020, 10.1007/BF01608497}{{\em Commun.
  Math. Phys.} {\bfseries 43} (1975) 199--220}.

\bibitem{Giddings:1992hh}
S.~B. Giddings, ``{Black holes and massive remnants},''
  \href{http://dx.doi.org/10.1103/PhysRevD.46.1347}{{\em Phys. Rev. D}
  {\bfseries 46} (1992) 1347--1352},
  \href{http://arxiv.org/abs/hep-th/9203059}{{\ttfamily arXiv:hep-th/9203059}}.

\bibitem{Hartle:1996rp}
J.~B. Hartle, ``{Generalized quantum theory in evaporating black hole
  space-times},'' in {\em {Symposium on Black Holes and Relativistic Stars
  (dedicated to memory of S. Chandrasekhar)}}, pp.~195--219.
\newblock 12, 1996.
\newblock \href{http://arxiv.org/abs/gr-qc/9705022}{{\ttfamily
  arXiv:gr-qc/9705022}}.

\bibitem{Maldacena:2001kr}
J.~M. Maldacena, ``{Eternal black holes in anti-de Sitter},''
  \href{http://dx.doi.org/10.1088/1126-6708/2003/04/021}{{\em JHEP} {\bfseries
  04} (2003) 021}, \href{http://arxiv.org/abs/hep-th/0106112}{{\ttfamily
  arXiv:hep-th/0106112}}.

\bibitem{Lunin:2001jy}
O.~Lunin and S.~D. Mathur, ``{AdS / CFT duality and the black hole information
  paradox},'' \href{http://dx.doi.org/10.1016/S0550-3213(01)00620-4}{{\em Nucl.
  Phys. B} {\bfseries 623} (2002) 342--394},
  \href{http://arxiv.org/abs/hep-th/0109154}{{\ttfamily arXiv:hep-th/0109154}}.

\bibitem{Horowitz:2003he}
G.~T. Horowitz and J.~M. Maldacena, ``{The Black hole final state},''
  \href{http://dx.doi.org/10.1088/1126-6708/2004/02/008}{{\em JHEP} {\bfseries
  02} (2004) 008}, \href{http://arxiv.org/abs/hep-th/0310281}{{\ttfamily
  arXiv:hep-th/0310281}}.

\bibitem{Mathur:2005zp}
S.~D. Mathur, ``{The Fuzzball proposal for black holes: An Elementary
  review},'' \href{http://dx.doi.org/10.1002/prop.200410203}{{\em Fortsch.
  Phys.} {\bfseries 53} (2005) 793--827},
  \href{http://arxiv.org/abs/hep-th/0502050}{{\ttfamily arXiv:hep-th/0502050}}.

\bibitem{Skenderis:2008qn}
K.~Skenderis and M.~Taylor, ``{The fuzzball proposal for black holes},''
  \href{http://dx.doi.org/10.1016/j.physrep.2008.08.001}{{\em Phys. Rept.}
  {\bfseries 467} (2008) 117--171},
  \href{http://arxiv.org/abs/0804.0552}{{\ttfamily arXiv:0804.0552 [hep-th]}}.

\bibitem{Almheiri:2012rt}
A.~Almheiri, D.~Marolf, J.~Polchinski, and J.~Sully, ``{Black Holes:
  Complementarity or Firewalls?},''
  \href{http://dx.doi.org/10.1007/JHEP02(2013)062}{{\em JHEP} {\bfseries 02}
  (2013) 062}, \href{http://arxiv.org/abs/1207.3123}{{\ttfamily arXiv:1207.3123
  [hep-th]}}.

\bibitem{Papadodimas:2012aq}
K.~Papadodimas and S.~Raju, ``{An Infalling Observer in AdS/CFT},''
  \href{http://dx.doi.org/10.1007/JHEP10(2013)212}{{\em JHEP} {\bfseries 10}
  (2013) 212}, \href{http://arxiv.org/abs/1211.6767}{{\ttfamily arXiv:1211.6767
  [hep-th]}}.

\bibitem{Almheiri:2013hfa}
A.~Almheiri, D.~Marolf, J.~Polchinski, D.~Stanford, and J.~Sully, ``{An
  Apologia for Firewalls},''
  \href{http://dx.doi.org/10.1007/JHEP09(2013)018}{{\em JHEP} {\bfseries 09}
  (2013) 018}, \href{http://arxiv.org/abs/1304.6483}{{\ttfamily arXiv:1304.6483
  [hep-th]}}.

\bibitem{Papadodimas:2013wnh}
K.~Papadodimas and S.~Raju, ``{Black Hole Interior in the Holographic
  Correspondence and the Information Paradox},''
  \href{http://dx.doi.org/10.1103/PhysRevLett.112.051301}{{\em Phys. Rev.
  Lett.} {\bfseries 112} no.~5, (2014) 051301},
  \href{http://arxiv.org/abs/1310.6334}{{\ttfamily arXiv:1310.6334 [hep-th]}}.

\bibitem{Papadodimas:2013jku}
K.~Papadodimas and S.~Raju, ``{State-Dependent Bulk-Boundary Maps and Black
  Hole Complementarity},''
  \href{http://dx.doi.org/10.1103/PhysRevD.89.086010}{{\em Phys. Rev. D}
  {\bfseries 89} no.~8, (2014) 086010},
  \href{http://arxiv.org/abs/1310.6335}{{\ttfamily arXiv:1310.6335 [hep-th]}}.

\bibitem{Bradler:2013gqa}
K.~Br\'adler and C.~Adami, ``{The capacity of black holes to transmit quantum
  information},'' \href{http://dx.doi.org/10.1007/JHEP05(2014)095}{{\em JHEP}
  {\bfseries 05} (2014) 095}, \href{http://arxiv.org/abs/1310.7914}{{\ttfamily
  arXiv:1310.7914 [quant-ph]}}.

\bibitem{Hawking:2016msc}
S.~W. Hawking, M.~J. Perry, and A.~Strominger, ``{Soft Hair on Black Holes},''
  \href{http://dx.doi.org/10.1103/PhysRevLett.116.231301}{{\em Phys. Rev.
  Lett.} {\bfseries 116} no.~23, (2016) 231301},
  \href{http://arxiv.org/abs/1601.00921}{{\ttfamily arXiv:1601.00921
  [hep-th]}}.

\bibitem{Penington:2019npb}
G.~Penington, ``{Entanglement Wedge Reconstruction and the Information
  Paradox},'' \href{http://dx.doi.org/10.1007/JHEP09(2020)002}{{\em JHEP}
  {\bfseries 09} (2020) 002}, \href{http://arxiv.org/abs/1905.08255}{{\ttfamily
  arXiv:1905.08255 [hep-th]}}.

\bibitem{Almheiri:2019psf}
A.~Almheiri, N.~Engelhardt, D.~Marolf, and H.~Maxfield, ``{The entropy of bulk
  quantum fields and the entanglement wedge of an evaporating black hole},''
  \href{http://dx.doi.org/10.1007/JHEP12(2019)063}{{\em JHEP} {\bfseries 12}
  (2019) 063}, \href{http://arxiv.org/abs/1905.08762}{{\ttfamily
  arXiv:1905.08762 [hep-th]}}.

\bibitem{Penington:2019kki}
G.~Penington, S.~H. Shenker, D.~Stanford, and Z.~Yang, ``{Replica wormholes and
  the black hole interior},''
  \href{http://dx.doi.org/10.1007/JHEP03(2022)205}{{\em JHEP} {\bfseries 03}
  (2022) 205}, \href{http://arxiv.org/abs/1911.11977}{{\ttfamily
  arXiv:1911.11977 [hep-th]}}.

\bibitem{Almheiri:2019qdq}
A.~Almheiri, T.~Hartman, J.~Maldacena, E.~Shaghoulian, and A.~Tajdini,
  ``{Replica Wormholes and the Entropy of Hawking Radiation},''
  \href{http://dx.doi.org/10.1007/JHEP05(2020)013}{{\em JHEP} {\bfseries 05}
  (2020) 013}, \href{http://arxiv.org/abs/1911.12333}{{\ttfamily
  arXiv:1911.12333 [hep-th]}}.

\bibitem{Preskill:1992tc}
J.~Preskill, ``{Do black holes destroy information?},'' in {\em {International
  Symposium on Black holes, Membranes, Wormholes and Superstrings}}.
\newblock 1, 1992.
\newblock \href{http://arxiv.org/abs/hep-th/9209058}{{\ttfamily
  arXiv:hep-th/9209058}}.

\bibitem{Giddings:1995gd}
S.~B. Giddings, ``{The Black hole information paradox},'' in {\em {PASCOS /
  HOPKINS 1995 (Joint Meeting of the International Symposium on Particles,
  Strings and Cosmology and the 19th Johns Hopkins Workshop on Current Problems
  in Particle Theory)}}, pp.~415--428.
\newblock 8, 1995.
\newblock \href{http://arxiv.org/abs/hep-th/9508151}{{\ttfamily
  arXiv:hep-th/9508151}}.

\bibitem{Mathur:2009hf}
S.~D. Mathur, ``{The Information paradox: A Pedagogical introduction},''
  \href{http://dx.doi.org/10.1088/0264-9381/26/22/224001}{{\em Class. Quant.
  Grav.} {\bfseries 26} (2009) 224001},
  \href{http://arxiv.org/abs/0909.1038}{{\ttfamily arXiv:0909.1038 [hep-th]}}.

\bibitem{Ashtekar:2020ifw}
A.~Ashtekar, ``{Black Hole evaporation: A Perspective from Loop Quantum
  Gravity},'' \href{http://dx.doi.org/10.3390/universe6020021}{{\em Universe}
  {\bfseries 6} no.~2, (2020) 21},
  \href{http://arxiv.org/abs/2001.08833}{{\ttfamily arXiv:2001.08833 [gr-qc]}}.

\bibitem{Almheiri:2020cfm}
A.~Almheiri, T.~Hartman, J.~Maldacena, E.~Shaghoulian, and A.~Tajdini, ``{The
  entropy of Hawking radiation},''
  \href{http://dx.doi.org/10.1103/RevModPhys.93.035002}{{\em Rev. Mod. Phys.}
  {\bfseries 93} no.~3, (2021) 035002},
  \href{http://arxiv.org/abs/2006.06872}{{\ttfamily arXiv:2006.06872
  [hep-th]}}.

\bibitem{Page:1993wv}
D.~N. Page, ``{Information in black hole radiation},''
  \href{http://dx.doi.org/10.1103/PhysRevLett.71.3743}{{\em Phys. Rev. Lett.}
  {\bfseries 71} (1993) 3743--3746},
\href{http://arxiv.org/abs/hep-th/9306083}{{\ttfamily arXiv:hep-th/9306083
  [hep-th]}}.

\bibitem{Page:2013dx}
D.~N. Page, ``{Time Dependence of Hawking Radiation Entropy},''
  \href{http://dx.doi.org/10.1088/1475-7516/2013/09/028}{{\em JCAP} {\bfseries
  1309} (2013) 028},
\href{http://arxiv.org/abs/1301.4995}{{\ttfamily arXiv:1301.4995 [hep-th]}}.

\bibitem{Kerr:1963ud}
R.~P. Kerr, ``{Gravitational field of a spinning mass as an example of
  algebraically special metrics},''
\href{http://dx.doi.org/10.1103/PhysRevLett.11.237}{{\em Phys. Rev. Lett.}
  {\bfseries 11} (1963) 237--238}.

\bibitem{LIGOScientific:2016aoc}
{\bfseries LIGO Scientific, Virgo} Collaboration, B.~P. Abbott {\em et~al.},
  ``{Observation of Gravitational Waves from a Binary Black Hole Merger},''
  \href{http://dx.doi.org/10.1103/PhysRevLett.116.061102}{{\em Phys. Rev.
  Lett.} {\bfseries 116} no.~6, (2016) 061102},
  \href{http://arxiv.org/abs/1602.03837}{{\ttfamily arXiv:1602.03837 [gr-qc]}}.

\bibitem{Nian:2019buz}
J.~Nian, ``{Kerr Black Hole Evaporation and Page Curve},''
  \href{http://arxiv.org/abs/1912.13474}{{\ttfamily arXiv:1912.13474
  [hep-th]}}.

\bibitem{MorrisThorne1988}
M.~S. Morris and K.~S. Thorne, ``{Wormholes in spacetime and their use for
  interstellar travel: A tool for teaching general relativity},''
  \href{http://dx.doi.org/10.1119/1.15620}{{\em American Journal of Physics}
  {\bfseries 56} no.~5, (05, 1988) 395--412}.

\bibitem{MorrisThorneYurtsever1988}
M.~S. Morris, K.~S. Thorne, and U.~Yurtsever, ``Wormholes, time machines, and
  the weak energy condition,''
  \href{http://dx.doi.org/10.1103/PhysRevLett.61.1446}{{\em Phys. Rev. Lett.}
  {\bfseries 61} (Sep, 1988) 1446--1449}.

\bibitem{Wald:1995yp}
R.~M. Wald, {\em {Quantum Field Theory in Curved Space-Time and Black Hole
  Thermodynamics}}.
\newblock Chicago Lectures in Physics. University of Chicago Press, Chicago,
  IL, 1995.

\bibitem{Hawking:1973uf}
S.~W. Hawking and G.~F.~R. Ellis,
  \href{http://dx.doi.org/10.1017/9781009253161}{{\em {The Large Scale
  Structure of Space-Time}}}.
\newblock Cambridge Monographs on Mathematical Physics. Cambridge University
  Press, 2, 2023.

\bibitem{Bekenstein:1997bt}
J.~D. Bekenstein, ``{Quantum black holes as atoms},'' in {\em {8th Marcel
  Grossmann Meeting on Recent Developments in Theoretical and Experimental
  General Relativity, Gravitation and Relativistic Field Theories (MG 8)}},
  pp.~92--111.
\newblock 6, 1997.
\newblock \href{http://arxiv.org/abs/gr-qc/9710076}{{\ttfamily
  arXiv:gr-qc/9710076}}.

\bibitem{EmparanSachs}
R.~Emparan and I.~Sachs, ``Quantization of ${\mathrm{ads}}_{3}$ black holes in
  external fields: Semiclassical results from pure gravity,''
  \href{http://dx.doi.org/10.1103/PhysRevLett.81.2408}{{\em Phys. Rev. Lett.}
  {\bfseries 81} (Sep, 1998) 2408--2411}.

\bibitem{Teukolsky1972}
S.~A. Teukolsky, ``Rotating black holes: Separable wave equations for
  gravitational and electromagnetic perturbations,''
  \href{http://dx.doi.org/10.1103/PhysRevLett.29.1114}{{\em Phys. Rev. Lett.}
  {\bfseries 29} (Oct, 1972) 1114--1118}.

\bibitem{Starobinsky:1973aij}
A.~A. Starobinsky, ``{Amplification of waves reflected from a rotating ''black
  hole''.},'' {\em Sov. Phys. JETP} {\bfseries 37} no.~1, (1973) 28--32.

\bibitem{Starobinsky:1974nkd}
A.~A. Starobinsky and S.~M. Churilov, ``{Amplification of electromagnetic and
  gravitational waves scattered by a rotating ''black hole''},'' {\em Sov.
  Phys. JETP} {\bfseries 65} no.~1, (1974) 1--5.

\bibitem{Teukolsky:1974yv}
S.~A. Teukolsky and W.~H. Press, ``{Perturbations of a rotating black hole. III
  - Interaction of the hole with gravitational and electromagnet ic
  radiation},'' \href{http://dx.doi.org/10.1086/153180}{{\em Astrophys. J.}
  {\bfseries 193} (1974) 443--461}.

\bibitem{Page1976}
D.~N. Page, ``Particle emission rates from a black hole. ii. massless particles
  from a rotating hole,''
  \href{http://dx.doi.org/10.1103/PhysRevD.14.3260}{{\em Phys. Rev. D}
  {\bfseries 14} (Dec, 1976) 3260--3273}.

\bibitem{Compere:2012jk}
G.~Comp\`ere, ``{The Kerr/CFT correspondence and its extensions},''
  \href{http://dx.doi.org/10.1007/s41114-017-0003-2}{{\em Living Rev. Rel.}
  {\bfseries 15} (2012) 11}, \href{http://arxiv.org/abs/1203.3561}{{\ttfamily
  arXiv:1203.3561 [hep-th]}}.

\bibitem{Guica:2008mu}
M.~Guica, T.~Hartman, W.~Song, and A.~Strominger, ``{The Kerr/CFT
  Correspondence},'' \href{http://dx.doi.org/10.1103/PhysRevD.80.124008}{{\em
  Phys. Rev.} {\bfseries D80} (2009) 124008},
\href{http://arxiv.org/abs/0809.4266}{{\ttfamily arXiv:0809.4266 [hep-th]}}.

\bibitem{Bardeen:1999px}
J.~M. Bardeen and G.~T. Horowitz, ``{The Extreme Kerr throat geometry: A Vacuum
  analog of AdS(2) x S**2},''
  \href{http://dx.doi.org/10.1103/PhysRevD.60.104030}{{\em Phys. Rev.}
  {\bfseries D60} (1999) 104030},
\href{http://arxiv.org/abs/hep-th/9905099}{{\ttfamily arXiv:hep-th/9905099
  [hep-th]}}.

\bibitem{Compere:2013bya}
G.~Comp\`ere, W.~Song, and A.~Strominger, ``{New Boundary Conditions for
  AdS3},'' \href{http://dx.doi.org/10.1007/JHEP05(2013)152}{{\em JHEP}
  {\bfseries 05} (2013) 152}, \href{http://arxiv.org/abs/1303.2662}{{\ttfamily
  arXiv:1303.2662 [hep-th]}}.

\bibitem{Lu:2008jk}
H.~Lu, J.~Mei, and C.~N. Pope, ``{Kerr/CFT Correspondence in Diverse
  Dimensions},'' \href{http://dx.doi.org/10.1088/1126-6708/2009/04/054}{{\em
  JHEP} {\bfseries 04} (2009) 054},
\href{http://arxiv.org/abs/0811.2225}{{\ttfamily arXiv:0811.2225 [hep-th]}}.

\bibitem{David:2020ems}
M.~David, J.~Nian, and L.~A. Pando~Zayas, ``{Gravitational Cardy Limit and AdS
  Black Hole Entropy},'' \href{http://dx.doi.org/10.1007/JHEP11(2020)041}{{\em
  JHEP} {\bfseries 11} (2020) 041},
  \href{http://arxiv.org/abs/2005.10251}{{\ttfamily arXiv:2005.10251
  [hep-th]}}.

\bibitem{Bredberg:2009pv}
I.~Bredberg, T.~Hartman, W.~Song, and A.~Strominger, ``{Black Hole
  Superradiance From Kerr/CFT},''
  \href{http://dx.doi.org/10.1007/JHEP04(2010)019}{{\em JHEP} {\bfseries 04}
  (2010) 019}, \href{http://arxiv.org/abs/0907.3477}{{\ttfamily arXiv:0907.3477
  [hep-th]}}.

\bibitem{Hartman:2009nz}
T.~Hartman, W.~Song, and A.~Strominger, ``{Holographic Derivation of
  Kerr-Newman Scattering Amplitudes for General Charge and Spin},''
  \href{http://dx.doi.org/10.1007/JHEP03(2010)118}{{\em JHEP} {\bfseries 03}
  (2010) 118}, \href{http://arxiv.org/abs/0908.3909}{{\ttfamily arXiv:0908.3909
  [hep-th]}}.

\bibitem{Chen:2010bh}
B.~Chen and J.~Long, ``{On Holographic description of the Kerr-Newman-AdS-dS
  black holes},'' \href{http://dx.doi.org/10.1007/JHEP08(2010)065}{{\em JHEP}
  {\bfseries 08} (2010) 065}, \href{http://arxiv.org/abs/1006.0157}{{\ttfamily
  arXiv:1006.0157 [hep-th]}}.

\bibitem{Nian:2020qsk}
J.~Nian and L.~A. Pando~Zayas, ``{Toward an effective CFT$_{2}$ from $
  \mathcal{N} $ = 4 super Yang-Mills and aspects of Hawking radiation},''
  \href{http://dx.doi.org/10.1007/JHEP07(2020)120}{{\em JHEP} {\bfseries 07}
  (2020) 120}, \href{http://arxiv.org/abs/2003.02770}{{\ttfamily
  arXiv:2003.02770 [hep-th]}}.

\bibitem{David:2020jhp}
M.~David and J.~Nian, ``{Universal entropy and hawking radiation of
  near-extremal AdS$_{4}$ black holes},''
  \href{http://dx.doi.org/10.1007/JHEP04(2021)256}{{\em JHEP} {\bfseries 04}
  (2021) 256}, \href{http://arxiv.org/abs/2009.12370}{{\ttfamily
  arXiv:2009.12370 [hep-th]}}.

\bibitem{Nian:2020bzf}
J.~Nian and L.~A.~P. Zayas, ``{Retarded Green's functions from rotating AdS
  black holes: Emergent CFT2 and viscosity},''
  \href{http://dx.doi.org/10.1103/PhysRevD.104.066020}{{\em Phys. Rev. D}
  {\bfseries 104} no.~6, (2021) 066020},
  \href{http://arxiv.org/abs/2012.02797}{{\ttfamily arXiv:2012.02797
  [hep-th]}}.

\bibitem{Castro:2010fd}
A.~Castro, A.~Maloney, and A.~Strominger, ``{Hidden Conformal Symmetry of the
  Kerr Black Hole},'' \href{http://dx.doi.org/10.1103/PhysRevD.82.024008}{{\em
  Phys. Rev. D} {\bfseries 82} (2010) 024008},
  \href{http://arxiv.org/abs/1004.0996}{{\ttfamily arXiv:1004.0996 [hep-th]}}.

\bibitem{Haco:2018ske}
S.~Haco, S.~W. Hawking, M.~J. Perry, and A.~Strominger, ``{Black Hole Entropy
  and Soft Hair},'' \href{http://dx.doi.org/10.1007/JHEP12(2018)098}{{\em JHEP}
  {\bfseries 12} (2018) 098}, \href{http://arxiv.org/abs/1810.01847}{{\ttfamily
  arXiv:1810.01847 [hep-th]}}.

\bibitem{Hawking:2016sgy}
S.~W. Hawking, M.~J. Perry, and A.~Strominger, ``{Superrotation Charge and
  Supertranslation Hair on Black Holes},''
  \href{http://dx.doi.org/10.1007/JHEP05(2017)161}{{\em JHEP} {\bfseries 05}
  (2017) 161}, \href{http://arxiv.org/abs/1611.09175}{{\ttfamily
  arXiv:1611.09175 [hep-th]}}.

\bibitem{Nian:2023dng}
J.~Nian and W.~Tian, ``{Gravitational Waves of Non-Extremal Kerr Black Holes
  from Conformal Symmetry},'' \href{http://arxiv.org/abs/2308.03577}{{\ttfamily
  arXiv:2308.03577 [hep-th]}}.

\bibitem{Maldacena:1997ih}
J.~M. Maldacena and A.~Strominger, ``{Universal low-energy dynamics for
  rotating black holes},''
  \href{http://dx.doi.org/10.1103/PhysRevD.56.4975}{{\em Phys. Rev. D}
  {\bfseries 56} (1997) 4975--4983},
  \href{http://arxiv.org/abs/hep-th/9702015}{{\ttfamily arXiv:hep-th/9702015}}.

\bibitem{Gubser:1997cm}
S.~S. Gubser, ``{Absorption of photons and fermions by black holes in
  four-dimensions},'' \href{http://dx.doi.org/10.1103/PhysRevD.56.7854}{{\em
  Phys. Rev. D} {\bfseries 56} (1997) 7854--7868},
  \href{http://arxiv.org/abs/hep-th/9706100}{{\ttfamily arXiv:hep-th/9706100}}.

\bibitem{Han:2023wxg}
M.~Han, C.~Rovelli, and F.~Soltani, ``{Geometry of the black-to-white hole
  transition within a single asymptotic region},''
  \href{http://dx.doi.org/10.1103/PhysRevD.107.064011}{{\em Phys. Rev. D}
  {\bfseries 107} no.~6, (2023) 064011},
  \href{http://arxiv.org/abs/2302.03872}{{\ttfamily arXiv:2302.03872 [gr-qc]}}.

\bibitem{Boyer:1966qh}
R.~H. Boyer and R.~W. Lindquist, ``{Maximal analytic extension of the Kerr
  metric},'' \href{http://dx.doi.org/10.1063/1.1705193}{{\em J. Math. Phys.}
  {\bfseries 8} (1967) 265}.

\bibitem{Roken:2015fja}
C.~R\"oken, ``{The massive Dirac equation in Kerr geometry: separability in
  Eddington\textendash{}Finkelstein-type coordinates and asymptotics},''
  \href{http://dx.doi.org/10.1007/s10714-017-2194-y}{{\em Gen. Rel. Grav.}
  {\bfseries 49} no.~3, (2017) 39},
  \href{http://arxiv.org/abs/1506.08038}{{\ttfamily arXiv:1506.08038 [gr-qc]}}.

\bibitem{Parikh:1999mf}
M.~K. Parikh and F.~Wilczek, ``{Hawking radiation as tunneling},''
  \href{http://dx.doi.org/10.1103/PhysRevLett.85.5042}{{\em Phys. Rev. Lett.}
  {\bfseries 85} (2000) 5042--5045},
\href{http://arxiv.org/abs/hep-th/9907001}{{\ttfamily arXiv:hep-th/9907001
  [hep-th]}}.

\bibitem{Wen:2019bjp}
W.-Y. Wen, ``{Hawking radiation as stimulated emission},''
  \href{http://dx.doi.org/10.1016/j.physletb.2020.135348}{{\em Phys. Lett. B}
  {\bfseries 803} (2020) 135348},
  \href{http://arxiv.org/abs/1910.06055}{{\ttfamily arXiv:1910.06055
  [hep-th]}}.

\bibitem{Zeng:2021kyb}
D.-f. Zeng, ``{Spontaneous radiation of black holes},''
  \href{http://dx.doi.org/10.1016/j.nuclphysb.2022.115722}{{\em Nucl. Phys. B}
  {\bfseries 977} (2022) 115722},
  \href{http://arxiv.org/abs/2112.12531}{{\ttfamily arXiv:2112.12531
  [hep-th]}}.

\bibitem{Zeng:2022brh}
D.-f. Zeng, ``{Gravity induced spontaneous radiation},''
  \href{http://dx.doi.org/10.1016/j.nuclphysb.2023.116171}{{\em Nucl. Phys. B}
  {\bfseries 990} (2023) 116171},
  \href{http://arxiv.org/abs/2207.05158}{{\ttfamily arXiv:2207.05158
  [hep-th]}}.

\bibitem{Chu:2022ieq}
C.-S. Chu and R.-X. Miao, ``{Tunneling of Bell Particles, Page Curve and Black
  Hole Information},'' \href{http://arxiv.org/abs/2209.03610}{{\ttfamily
  arXiv:2209.03610 [hep-th]}}.

\bibitem{Chu:2023agv}
C.-S. Chu and R.-X. Miao, ``{Tunneling, Page Curve and Black Hole
  Information},'' \href{http://arxiv.org/abs/2307.06176}{{\ttfamily
  arXiv:2307.06176 [hep-th]}}.

\bibitem{Feng:2023pfq}
Z.-W. Feng, Y.~Ling, X.-N. Wu, and Q.-Q. Jiang, ``{New black-to-white hole
  solutions with improved geometry and energy conditions},''
  \href{http://arxiv.org/abs/2308.15689}{{\ttfamily arXiv:2308.15689 [gr-qc]}}.

\bibitem{QuInfoBook}
M.~Nielsen and I.~Chuang, {\em Quantum Computation and Quantum Information}.
\newblock Cambridge Series on Information and the Natural Sciences. Cambridge
  University Press, 2000.

\bibitem{Ho:2022gpg}
P.-M. Ho and H.~Kawai, ``{UV and IR effects on Hawking radiation},''
  \href{http://dx.doi.org/10.1007/JHEP03(2023)002}{{\em JHEP} {\bfseries 03}
  (2023) 002}, \href{http://arxiv.org/abs/2207.07122}{{\ttfamily
  arXiv:2207.07122 [hep-th]}}.

\bibitem{McNamara}
J.~M. McNamara, ``Instability of black hole inner horizons,'' {\em Proceedings
  of the Royal Society of London. Series A, Mathematical and Physical Sciences}
  {\bfseries 358} no.~1695, (1978) 499--517.

\bibitem{Penrose}
R.~{Penrose}, ``{Singularities of spacetime.},'' in {\em Theoretical Principles
  in Astrophysics and Relativity}, N.~R. {Lebovitz}, ed., pp.~217--243.
\newblock 1978.

\bibitem{Dafermos:2017dbw}
M.~Dafermos and J.~Luk, ``{The interior of dynamical vacuum black holes I: The
  $C^0$-stability of the Kerr Cauchy horizon},''
  \href{http://arxiv.org/abs/1710.01722}{{\ttfamily arXiv:1710.01722 [gr-qc]}}.

\bibitem{Dias:2018etb}
O.~J.~C. Dias, H.~S. Reall, and J.~E. Santos, ``{Strong cosmic censorship:
  taking the rough with the smooth},''
  \href{http://dx.doi.org/10.1007/JHEP10(2018)001}{{\em JHEP} {\bfseries 10}
  (2018) 001}, \href{http://arxiv.org/abs/1808.02895}{{\ttfamily
  arXiv:1808.02895 [gr-qc]}}.

\bibitem{Dray:1984ha}
T.~Dray and G.~'t~Hooft, ``{The Gravitational Shock Wave of a Massless
  Particle},'' \href{http://dx.doi.org/10.1016/0550-3213(85)90525-5}{{\em Nucl.
  Phys. B} {\bfseries 253} (1985) 173--188}.

\bibitem{Gehrels:2006tk}
N.~Gehrels {\em et~al.}, ``{Swift detects a remarkable gamma-ray burst, GRB
  060614, that introduces a new classification scheme},''
  \href{http://dx.doi.org/10.1038/nature05376}{{\em Nature} {\bfseries 444}
  (2006) 1044}, \href{http://arxiv.org/abs/astro-ph/0610635}{{\ttfamily
  arXiv:astro-ph/0610635}}.

\bibitem{BHexplosion}
S.~W. {Hawking}, ``{Black hole explosions?},''
  \href{http://dx.doi.org/10.1038/248030a0}{{\em {Nature}} {\bfseries 248}
  no.~5443, (Mar., 1974) 30--31}.

\end{thebibliography}\endgroup
\end{document}